\documentclass[prd,preprintnumbers,amsmath,amssymb,floatfix]{revtex4}
\usepackage{graphicx,subfig}
\usepackage{bm}
\usepackage{amsfonts}
\usepackage{lineno,hyperref}
\usepackage{array}
\usepackage{color,multirow,cases, makecell}
\usepackage{pifont, float, setspace, moresize, soul}
\usepackage{cancel}
\soulregister\ref7 
\soulregister\eqref7 
\soulregister\citealt7 

\begin{document}

\title{Stability Analysis of Cosmological Perturbations in the Bumblebee Model: Parameter Constraints and Gravitational Waves}

\author{Xiao-Bin Lai$^{a}$$^{b}$\footnote{laixb2024@lzu.edu.cn}}
\author{Yu-Qi Dong$^{a}$$^{b}$\footnote{dongyq2023@lzu.edu.cn}}
\author{Yu-Zhi Fan$^{a}$$^{b}$\footnote{fanyzh2025@lzu.edu.cn}}
\author{Yu-Xiao Liu$^{a}$$^{b}$\footnote{liuyx@lzu.edu.cn, corresponding author}}

\affiliation{
$^{a}$Lanzhou Center for Theoretical Physics,
  Key Laboratory of Theoretical Physics of Gansu Province, 
  Key Laboratory of Quantum Theory and Applications of MoE, 
  Gansu Provincial Research Center for Basic Disciplines of Quantum Physics,
  Lanzhou University, Lanzhou 730000, China \\
$^{b}$Institute of Theoretical Physics $\&$ Research Center of Gravitation,
  School of Physical Science and Technology,
  Lanzhou University, Lanzhou 730000, China
}

\begin{abstract}
  We constrain the parameter space of the Bumblebee model in a cosmological background and then investigate the properties of gravitational waves within the constrained parameter space. Our analysis reveals seven perturbative degrees of freedom in the cosmological background: two tensor, two vector, and two scalar modes, along with an additional mode from the matter sector. The stability conditions for all these modes are derived. By incorporating the observed accelerated expansion of the universe and the observational constraints on tensor gravitational waves, we derive bounds on the parameter space of the Bumblebee model. Our results indicate that the non-minimal coupling parameter $\xi$ must be non-positive, a constant background value $b_{t}$ of the Bumblebee field implies $\sigma\ne -\tfrac{1}{2}\xi$, and the Lorentz-violating parameter $\xi b^2$ has a lower bound on the order of $10^{-15}$. We then investigate the propagation characteristics and polarization modes of gravitational waves in both the small-scale and Minkowski limits. The propagation modes of gravitational waves in the Bumblebee model consist of two tensor modes, two vector modes, and one scalar mode. Notably, the tensor modes travel at subluminal speeds, whereas the vector and scalar modes propagate at superluminal speeds, when $\xi b_{t}^2\ne 0$. These results provide a concrete theoretical framework and specific observational signatures for testing Lorentz invariance in the gravitational sector with future gravitational-wave detectors.
\end{abstract}

\maketitle
\tableofcontents 

\section{Introduction}
Since its proposal, general relativity (GR) has undergone rigorous tests in both weak-field and strong-field regimes. In the weak-field limit, its predictions have been precisely verified through observations and experiments, such as the precession of Mercury's perihelion~\cite{einstein1915erklarung}, the deflection of light~\cite{dyson1923determination}, and the Pound-Rebka experiment~\cite{PhysRevLett.3.439,Pound:1960zz}. In the strong-field limit, the orbital decay of the Hulse-Taylor pulsar shows excellent agreement with the gravitational wave (GW) radiation predicted by GR~\cite{Hulse:1974eb,Taylor:1979zz}. The first direct detection of GWs, GW150914, confirms a key prediction of GR~\cite{LIGOScientific:2016aoc,LIGOScientific:2016emj}. Imaging of the black hole shadows of M87* and Sagittarius A* by the Event Horizon Telescope provides further support for the black hole solutions predicted by GR~\cite{EventHorizonTelescope:2019dse,EventHorizonTelescope:2022wkp,EventHorizonTelescope:2022xqj}. This series of observations and experiments has tested the validity of GR across both the weak-field regimes and the highly non-linear strong-field regimes.
\par
Although GR has successfully withstood many experimental tests, several critical issues remain difficult to explain within its framework, such as the dark matter problem~\cite{Smith:1936mlg,Zwicky:1937zza}, the dark energy problem~\cite{Peebles:2002gy}, the problem of quantization~\cite{tHooft:1974toh,Goroff:1985th}, and the hierarchy problem~\cite{Arkani-Hamed:1998jmv,Randall:1999ee,Randall:1999vf}. This has led to questions within the physics community about whether GR is the ultimate theory of gravity, which has spurred research into modified gravity theories.
\par
The main approaches for constructing modified gravity theories include~\cite{Nojiri:2010wj,Clifton:2011jh}:
\begin{itemize}
 \item[(1)] adding new fields: Brans-Dicke theory~\cite{Brans:1961sx}, Horndeski theory~\cite{Horndeski:1974wa}, Bumblebee theory~\cite{Kostelecky:2003fs}, and so on~\cite{Bekenstein:2004ne,Jacobson:2007veq,Heisenberg:2014rta,Geng:2015kvs}; 
 \item[(2)] considering higher-order derivatives: $f(R)$ theory~\cite{Sotiriou:2008rp}, $f(\mathcal{G})$ theory~\cite{DeFelice:2008wz}, and so on~\cite{Horava:2009uw,Helpin:2019kcq}; 
 \item[(3)] exploring higher dimensions: Kaluza-Klein theory~\cite{Kaluza:1921tu}, Randall-Sundrum theory~\cite{Randall:1999ee,Randall:1999vf}, and so on~\cite{Rubakov:1983bb,Arkani-Hamed:1998jmv}; 
 \item[(4)] modifying geometry: Palatini-$f(R)$ theory~\cite{buchdahl1970non}, $f(Q)$ theory~\cite{Heisenberg:2023lru}, and so on~\cite{Hehl:1994ue,Lu:2020eux,BeltranJimenez:2019esp}; 
 \item[(5)] others: non-local theory~\cite{Horava:2008ih,Deser:2007jk,Deffayet:2009ca}, spatially covariant gravity~\cite{Gleyzes:2013ooa,Gao:2014soa,Gao:2014fra}. 
\end{itemize}
Many of these modified gravity theories offer unique interpretations of GWs and the universe that differ from the interpretations provided by GR. For example, in general metric theories, up to six independent polarization modes of GWs are possible, significantly exceeding the two tensor modes predicted by GR~\cite{Eardley:1973zuo}. Certain modified gravity theories can effectively describe the evolution of the universe and provide a theoretical basis for explaining dark energy~\cite{Nojiri:2006ri,Nojiri:2017ncd}. Some modified gravity theories fit galaxy rotation curves well, offering viable alternatives to the dark matter hypothesis~\cite{Famaey:2011kh,Capozziello:2012ie}. Therefore, theoretical consistency analyses and experimental tests are crucial for identifying a more complete description of gravity.
\par
A theoretically consistent modified gravity theory must be stable. A critical issue in this regard is the ghost problem. If ghost degrees of freedom, i.e., field modes carrying negative energy, are present, the theory will inevitably lead to vacuum instability or even catastrophic decay~\cite{Woodard:2006nt}. Thus, the ghost-free condition serves as an important tool for selecting modified gravity theories and constraining theoretical parameters~\cite{DeFelice:2016yws,Kase:2018aps,Clough:2022ygm}. Especially, the ghost problem is almost ubiquitous in higher-order modified gravity theories. The well-known Ostrogradsky theorem~\cite{Ostrogradsky:1850fid} states that when a Lagrangian contains higher-order (second order or higher) time  derivatives of dynamical variables, its Hamiltonian is usually bounded neither from above nor from below~\cite{Woodard:2006nt}. This implies that positive and negative parts could be excited to arbitrarily high energies, leading to vacuum instability~\cite{Woodard:2006nt}. There are also other studies concerning theoretical consistency, such as the initial value problem and the stability of evolution~\cite{kreiss1989initial,Papallo:2017qvl,Bluhm:2008yt,Clough:2022ygm}. 
\par
For the experimental tests of modified gravity theories, one of the most direct and promising approaches is the detection and characterization of GW signals. The polarization modes and propagation speeds of GWs are two critically important characteristics that typically vary across different modified gravity theories, making them direct probes for testing gravity theories. In GR, GWs propagate at the speed of light and exhibit only two tensor polarization modes: the plus mode ($P_{+}$) and the cross mode ($P_{\times}$). However, a general metric theory in four-dimensional spacetime allows for GWs up to six independent polarization modes (the additional ones being breathing ($P_{b}$), longitudinal ($P_{l}$), vector-$x$ ($P_{x}$), and vector-$y$ ($P_{y}$) modes) to propagate at different speeds~\cite{Eardley:1973zuo}. Furthermore, within a torsionless framework of metric-affine theory, up to eight independent polarization modes are possible~\cite{Dong:2025ddi,Dong:2025pyz}, including additional shear-$x$ and shear-$y$ modes. The first multi-messenger observation, a binary neutron star coalescence GW170817~\cite{LIGOScientific:2017vwq} and the associated gamma-ray burst GRB170817A~\cite{Goldstein:2017mmi}, places a stringent constraint on the speed of tensor modes $c_{t}$: $-3\times 10^{-15} \le c_{t}-1 \le 7\times 10^{-16}$~\cite{Savchenko:2017ffs}. Clearly, its potential deviation from the speed of light is exceedingly minute. Considering the expansion of the universe and potential emission delays, it is now generally accepted that the propagation speed of tensor modes is equal to the speed of light. Numerous studies have examined GWs from an observational perspective, including imposing experimental constraints on theoretical models~\cite{Baker:2017hug,Creminelli:2017sry,Dong:2022cvf}, investigating their properties~\cite{Liang:2022hxd,Lai:2024fza,Dong:2024zal,Fan:2024pex,Wang:2025hla,Dong:2025ddi}, and others~\cite{Amarilo:2023wpn,Jia:2024pdk,DeFelice:2025ykh}.
\par
In this paper, we focus on the cosmological perturbations for the Bumblebee model with a perfect fluid. First, using the background equations and the condition of an accelerated expanding universe, we perform a preliminary analysis of the parameter space and provide a brief discussion on dark energy. Then, we consider the scalar perturbations, vector perturbations, and tensor perturbations separately. We begin by deriving the second-order perturbation actions. Subsequently, based on the invariance of the linearized theory under infinitesimal coordinate transformations, we choose appropriate gauge conditions. We then eliminate the non-dynamical variables in Fourier space, which yields an effective action for the dynamical variables only. Starting from the effective action, we first derive the ghost-free conditions for the perturbations and then, in the small-scale limit, analyze the stability and propagation characteristics of GWs, which further constrain the parameter space. Finally, within the constrained parameter space, we analyze the polarization modes, the number of degrees of freedom, the propagation speeds, and the amplitude relations among the different GW modes.
\par
This paper is structured as follows. In Sec.~\ref{sec2-9132322}, we briefly introduce the Bumblebee model and the Schutz-Sorkin action for a perfect fluid. In Sec.~\ref{Per-cosBack-1701}, we perform a scalar-vector-tensor decomposition of the variables, analyze the background field equations, and discuss dark energy. In Sec.~\ref{tensor-9132349}, we consider the tensor perturbations. We derive the second-order effective action, analyze the stability of the tensor perturbations, and constrain the Lorentz-violating parameter through GW observations. In Sec.~\ref{vector-section-9132359}, we derive the second-order effective action for the vector perturbations and constrain the parameter space based on the stability conditions. In Sec.~\ref{scalar-section-91409}, we focus on the scalar perturbations. For three different gauge choices, we derive the stability conditions, analyze the propagation characteristics of GWs, and constrain the parameter space. In Sec.~\ref{polarization-section-914013}, we analyze the polarization modes, the propagation speeds, and the amplitude relations of GWs within the constrained parameter space. Finally, we present our conclusion in Sec.~\ref{conclusion-section-914018}.
\par
Throughout this work, we restrict our analysis to four-dimensional spacetime. We adopt the following conventions: Greek indices ($\mu,\nu,\alpha,\beta,\dots$) denote spacetime coordinates, while Latin indices ($i,j,k,\dots$) denote spatial coordinates. The metric signature is taken to be $(-,+,+,+)$, and we work in units where the speed of light is $c = 1$.

\section{The Bumblebee model with a perfect fluid} \label{sec2-9132322}
The Bumblebee model~\cite{Kostelecky:2003fs} is a Lorentz-violating gravity theory that has a very simple form but encompasses interesting features, including rotation, boost, and CPT violations. Owing to its unique implications for our understanding of the universe, this model has been the subject of extensive research in GW physics and cosmology. It has been widely studied in the contexts of GWs~\cite{Liang:2022hxd}, black holes~\cite{Liu:2024axg,Filho:2022yrk}, and cosmology~\cite{vandeBruck:2025aaa}.
\par
The action of the Bumblebee model is given by~\cite{Kostelecky:2003fs}
\begin{eqnarray}
  S=\int d^4x\sqrt{-g}\left[ \frac{1}{2\kappa}(R-2\Lambda+\xi B^{\mu}B^{\nu}R_{\mu\nu}+\sigma B^{\mu}B_{\mu}R)-\frac{1}{4}B_{\mu\nu}B^{\mu\nu}-V(B^{\mu}B_{\mu}\pm b^2) \right]+S_{m}. \label{Bumblebee-action-1503}
\end{eqnarray}
Here, $\kappa=8\pi G$, $b^2$ is a real positive constant, $B_{\mu\nu}=\nabla_{\mu}B_{\nu}-\nabla_{\nu}B_{\mu}$ is the field strength of the Bumblebee field $B_{\mu}$, and $\xi$ and $\sigma$ are real coupling constants. The sign in $V(B^{\mu}B_{\mu}\pm b^2)$ depends on whether $B_{\mu}$ is timelike or spacelike. The Bumblebee field gives rise to Lorentz violation through the potential $V$, which provides a nonzero vacuum expectation value $b_{\mu}$ for $B_{\mu}$. In this paper, we consider a spatially isotropic cosmological background. Thus, we take $b_{\mu}=(b_{t},0,0,0)$ with $b_{t}^2=b^2$.
\par
In its rest frame, a perfect fluid is uniquely characterized by its energy density and pressure. For a perfect fluid that does not couple explicitly to the curvature, it is natural to choose either the energy density $\rho$ ($\mathcal{L}_m=-\rho$)~\cite{Brown:1992kc, hawking2023large} or the pressure $p$ ($\mathcal{L}_m=p$)~\cite{Brown:1992kc, Schutz:1970my} as the matter Lagrangian. Another suitable choice is $\mathcal{L}_m=-na$~\cite{Brown:1992kc, Bertolami:2008ab}, where $n$ is the particle number density and $a$ is the physical free energy per particle, defined by $a=\rho/n-Ts$, with $T$ denoting the temperature and $s$ the entropy per particle. The three Lagrangian densities are equivalent within the framework of GR~\cite{Brown:1992kc}. When matter couples nonminimally to the Ricci scalar, several studies have been conducted~\cite{Bertolami:2008ab, Faraoni:2009rk}. For more related research on perfect fluids, see Refs.~\cite{deBoer:2017ing, Ovalle:2017fgl, Buchert:2001sa}.
\par
In this paper, we consider a perfect fluid that is minimally coupled to gravity and can be described by the Schutz-Sorkin action~\cite{Schutz:1977df, DeFelice:2009bx, Bertolami:2008ab, DeFelice:2016yws, Kase:2018nwt}
\begin{eqnarray}
  S_m=-\int d^4x \left[ \sqrt{-g}\rho_m(n)+J^{\mu}(\partial_{\mu}\ell+\mathcal{A}_1\partial_{\mu}\mathcal{B}_1+\mathcal{A}_2\partial_{\mu}\mathcal{B}_2) \right]. \label{perfect-fluid-action}
\end{eqnarray}
Here, $\rho_{m}$ is the energy density, $n$ the particle number density, $J^{\mu}$ a vector density, and $\ell$ a scalar. The quantities $\mathcal{A}_1$, $\mathcal{A}_2$, $\mathcal{B}_1$, and $\mathcal{B}_2$ are related to the intrinsic vector perturbations of the matter (see Refs.~\cite{DeFelice:2009bx, DeFelice:2016yws}).
\par 
Note that the action $S_m$ is a functional of $g_{\mu\nu}$, $J^{\mu}$, $\ell$, $\mathcal{A}_1$, $\mathcal{A}_2$, $\mathcal{B}_1$, and $\mathcal{B}_2$, i.e., $S_m=S_m[g_{\mu\nu},J^{\mu},\ell,\mathcal{A}_1,\mathcal{A}_2,\mathcal{B}_1,\mathcal{B}_2]$. The scalar field $\ell$ serves as a Lagrange multiplier enforcing the constraint $\partial_{\mu}J^{\mu}=0$, which expresses particle number conservation. The vector density $J^{\mu}$, which represents the particle number flux vector, is defined in terms of the number density $n$ and the four-velocity $U^\mu$ as
\begin{eqnarray}
    J^{\mu}=\sqrt{-g}nU^{\mu}.\label{vectorDensity-all}
\end{eqnarray}
The four-velocity $U^{\mu}$ satisfies the normalization condition $U^{\mu}U_{\mu}=-1$. Thus, the particle number density is given by $n=|J|/\sqrt{-g}$, and the energy density is consequently a function $\rho_{m}=\rho_{m}(|J|/\sqrt{-g})$. 
\par
Varying the action~\eqref{perfect-fluid-action} with respect to the metric $g_{\mu\nu}$, we obtain the energy-momentum tensor for a perfect fluid
\begin{eqnarray}
    T^{\mu\nu}=\rho_m U^{\mu}U^{\nu}+\left( n\frac{\partial\rho_m}{\partial n}-\rho_m \right)\left(g^{\mu\nu}+U^{\mu}U^{\nu}\right).
\end{eqnarray}
Here, we use the definition of the matter energy-momentum tensor $T_{\mu\nu}=-\frac{2}{\sqrt{-g}}\frac{\delta (\sqrt{-g}\mathcal{L}_m)}{\delta (g^{\mu\nu})}$. Now, consider the energy-momentum tensor of a perfect fluid, $T^{\mu\nu}=(\rho_m +p_m)U^{\mu}U^{\nu}+p_mg^{\mu\nu}$. By comparing these two expressions, the definition of pressure can be identified as
\begin{eqnarray}
    p_m=n\frac{\partial\rho_m}{\partial n}-\rho_m. \label{pm-2127}
\end{eqnarray}
\par
Since the gravitational action $S_g$ is independent of $J^{\mu}$, varying the total action~\eqref{Bumblebee-action-1503} with respect to the vector density $J^{\mu}$ yields
\begin{eqnarray}
  U_{\mu}\equiv \frac{J_{\mu}}{|J|}=\frac{1}{\rho_{m,n}}\left(\partial_{\mu}\ell+\mathcal{A}_1\partial_{\mu}\mathcal{B}_1+\mathcal{A}_2\partial_{\mu}\mathcal{B}_2\right),
\end{eqnarray}
where $\rho_{m,n}=\partial\rho_m/\partial n$. One can show that the spatial components $U_i$ of $U_{\mu}$ can be decomposed into a scalar part and a divergence-free vector part. This decomposition holds even if $\rho_{m,n}$ is constant, which is consistent with Refs.~\cite{DeFelice:2009bx, Schutz:1970my}. In the cosmological background, the divergence-free vector part of $U_i$ is generated by the scalars $\mathcal{A}_1$, $\mathcal{A}_2$, $\mathcal{B}_1$, and $\mathcal{B}_2$.  

\section{Perturbations and cosmological background}\label{Per-cosBack-1701}
In this paper, we focus on the Bumblebee model~\eqref{Bumblebee-action-1503} with a perfect fluid described by the Schutz-Sorkin action~\eqref{perfect-fluid-action}. The total action is
\begin{eqnarray}
  S&=&\int d^4x\sqrt{-g}\left[ \frac{1}{2\kappa}\left(R-2\Lambda+\xi B^{\mu}B^{\nu}R_{\mu\nu}+\sigma B^{\mu}B_{\mu}R\right)-\frac{1}{4}B_{\mu\nu}B^{\mu\nu}-V(B^{\mu}B_{\mu}\pm b^2) \right] \nonumber
  \\
  &&-\int d^4x \left[ \sqrt{-g}\rho_m(n)+J^{\mu}\left(\partial_{\mu}\ell+\mathcal{A}_1\partial_{\mu}\mathcal{B}_1+\mathcal{A}_2\partial_{\mu}\mathcal{B}_2\right) \right].\label{action-all}
\end{eqnarray}
Observations have shown that the current universe is very close to a spatially flat geometry~\cite{WMAP:2006bqn}. Therefore, we will analyze the equations of motion for the Bumblebee model within a spatially flat cosmological background
\begin{eqnarray}
    ds^2=-dt^2+a^2(t)\delta_{ij}dx^{i}dx^{j}.\label{FRW-metric}
\end{eqnarray}
Here, $a(t)$ is the scale factor. The universe described by this metric is spatially homogeneous and isotropic. It is therefore natural to choose the background fields
\begin{eqnarray}
    &&\overline{B}_{\mu}=(b_{t},0,0,0),\label{BG-vector}\\
    &&\overline{J}^{\mu}=\big(\overline{J},0,0,0\big).
\end{eqnarray}
Here and in what follows, the symbol ``$\overline{\;\;}$" above the physical quantities denotes the corresponding background quantities, and $b_{t}$ is a constant. Especially, in a comoving coordinate system, $\overline{J}$ is a constant, see Eq.~\eqref{matter-partial-t}.

\subsection{Perturbations in a cosmological background}\label{Pertur-FRW-subsec}
For a spatially homogeneous and isotropic universe, field perturbations can always be decomposed into spatial tensor, vector, and scalar parts, according to the transformation properties under 3-dimensional spatial rotation. This decomposition has been introduced in Refs.~\cite{Flanagan:2005yc, Jackiw:2003pm}. Using this, the metric $g_{\mu\nu}$, the Bumblebee field $B_{\mu}$, the vector density $J^{\mu}$, and the scalar $\ell$ in a cosmological background can be decomposed as
\begin{eqnarray}
  &&ds^2=-(1+2\phi_{h})dt^2+2(\lambda_{i}+\partial_{i}\varphi_{h})dx^{i}dt+a^2\left[ \delta_{ij}+h^{\mathrm{TT}}_{ij}+2\partial_{(i}\varepsilon_{j)}+E\delta_{ij}+\partial_{i}\partial_{j}\alpha \right]dx^{i}dx^{j},\label{ds-full-1514}
  \\
  &&B_{\mu}=\overline{B}_{\mu}+(\phi_{b},\; \zeta_{i}+\partial_{i}\varphi_{b}),
  \\
  &&J^{\mu}=\overline{J}^{\mu}+\left(\phi_{m},\; \chi^{i}+\frac{1}{a^2}\delta^{ij}\partial_{j}\varphi_{m}\right), \label{J-full-1514}
  \\
  &&\ell=\overline{\ell}(t)+\phi_{\ell}. \label{ell-full-1514}
\end{eqnarray}
Here, $h^{\mathrm{TT}}_{ij}$ is a transverse-traceless tensor and $\lambda_{i},\varepsilon_{i},\zeta_{i},\chi^{i}$ are transverse vectors, i.e.,
\begin{eqnarray}
  &&\partial^{i}h^{\mathrm{TT}}_{ij}=0,\quad \delta^{ij}h^{\mathrm{TT}}_{ij}=0,
  \\
  &&\partial^{i}\lambda_{i}=\partial^{i}\varepsilon_{i}=\partial^{i}\zeta_{i}=\partial_{i}\chi^{i}=0,
\end{eqnarray}
where $\partial^{i}=\delta^{ij}\partial_{j}$. The background quantity $\overline{\ell}$ is a function of $t$ only, as will be shown in Eq.~\eqref{ellBk-1614}. Note that the tensor perturbation $(h^{\mathrm{TT}}_{ij})$, the vector perturbations $(\lambda_{i},\varepsilon_{i},\zeta_{i},\chi^{i})$, and the scalar perturbations $(\phi_{h},\varphi_{h},E,\alpha,\phi_{b},\varphi_{b},\phi_{m},\varphi_{m},\phi_{\ell})$ are functions of the coordinates $t$, $x$, $y$, and $z$. Although $J^{\mu}$ is a vector density, the decomposition in Eq.~\eqref{J-full-1514} remains valid. This is because the first-order perturbation of $\sqrt{-g}$ vanishes, and the background quantity $\sqrt{-\overline{g}}$ is a function of $t$ only.
\par
For $\mathcal{A}_1$, $\mathcal{A}_2$, $\mathcal{B}_1$, and $\mathcal{B}_2$, we make the simplest choice that contains all the necessary information about the vector perturbations of matter~\cite{DeFelice:2009bx, DeFelice:2016yws}
\begin{eqnarray}
  \mathcal{A}_1=\delta\mathcal{A}_1(t,z),\quad \mathcal{A}_2=\delta\mathcal{A}_2(t,z),\quad \mathcal{B}_1=x+\delta\mathcal{B}_1(t,z),\quad \mathcal{B}_2=y+\delta\mathcal{B}_2(t,z). \label{AB-Perturbations-1117}
\end{eqnarray}
The quantities $\delta\mathcal{A}_1$, $\delta\mathcal{A}_2$, $\delta\mathcal{B}_1$, and $\delta\mathcal{B}_2$ are perturbations that depend on $t$ and $z$. Here, we work in a coordinate system where GWs propagate along the ``$+z$'' direction. Note that $\delta\mathcal{A}_{1,2}$ and $\delta\mathcal{B}_{1,2}$ contribute only to the vector perturbations of matter.
\par
In this theory, the tensor, vector, and scalar perturbations decouple from each other in the cosmological background. This allows us to study each sector independently, which greatly simplifies the subsequent analysis and calculations.

\subsection{Background equations} \label{BEofM}
First, we focus on the matter action~\eqref{perfect-fluid-action}. From Eq.~\eqref{vectorDensity-all}, the background value $\overline{J}^{\mu}$ of the vector density $J^{\mu}$ is obtained as
\begin{eqnarray}
    \overline{J}^{\mu}=(\overline{n} a^3,0,0,0).\label{JBackground-2124}
\end{eqnarray}
Here, we have used the value of the four-velocity in the comoving coordinates: $U^{\mu}=(1,0,0,0)$. Varying the action~\eqref{perfect-fluid-action} with respect to $J^{\mu}$ yields a constraint on $\overline{\ell}$
\begin{eqnarray}
  \dot{\overline{\ell}}=-\overline{\rho}_{m,n}, \label{ellBk-1614}
\end{eqnarray}
where $\partial_i\overline{\ell}=0$ has been omitted, which implies $\overline{\ell}$ is a function of $t$ only. Here and in what follows, a dot over a quantity denotes its time derivative, e.g., $\dot{\overline{n}}=\partial \overline{n}/\partial t$. Furthermore, the conservation of particle number, which follows from varying the matter action~\eqref{perfect-fluid-action} with respect to $\overline{\ell}$, yields
\begin{eqnarray}
    0&=&\partial_{\mu}\overline{J}^{\mu}=\partial_{t}(\overline{n} a^3)=\frac{\partial\overline{\rho}_m}{\partial \overline{n}}\dot{\overline{n}} a^3 +3\overline{n} \frac{\partial\overline{\rho}_m}{\partial \overline{n}} a^2 \dot{a} \nonumber\\
    &=&\dot{\overline{\rho}}_m+3H(\overline{\rho}_m+\overline{p}_m).\label{matter-partial-t}
\end{eqnarray}
Here, we have used the definition of the Hubble parameter, $H = \dot{a}/a$, and have multiplied the right-hand side of the third equal sign by ${\partial\overline{\rho}_m}/{\partial\overline{n}}$. Since the left-hand side of the equation is zero, this operation is valid.
\par
For ordinary matter, the energy density $\overline{\rho}_m$ is positive and corresponds to a non-negative pressure $\overline{p}_m$. Equation~\eqref{matter-partial-t} shows that for a static universe ($H = 0$), the energy density $\overline{\rho}_m$ remains constant. However, observational evidence indicates that the present universe is not only expanding but also accelerating~\cite{SupernovaCosmologyProject:1998vns, SupernovaSearchTeam:1998fmf, Riess:1998dv}. For an expanding universe, $H(t_0) > 0$ (where $t_0$ is the current time), which implies $\dot{\overline{\rho}}_m < 0$. Consequently, the energy density of ordinary matter decreases with expansion, which is a natural and physically intuitive result.
\par
To derive the Friedmann equation, we include the lapse function $N(t)$ in the metric~\eqref{FRW-metric}
\begin{eqnarray}
    ds^2=-N^2(t)dt^2+a^2(t)\delta_{ij}dx^{i}dx^{j}.\label{metric-N}
\end{eqnarray}
We finally set $N=1$ after varying the action~\eqref{action-all}. In this background, the Schutz-Sorkin action~\eqref{perfect-fluid-action} reduces to
\begin{eqnarray}
    \overline{S}_m=-\int d^4x a^3 \left(N\overline{\rho}_m+\overline{n}\partial_t \overline{\ell}\right).
\end{eqnarray}
Next, we substitute the background metric~\eqref{metric-N} and the background vector field~\eqref{BG-vector} into the full action~\eqref{action-all} to obtain the background action $\overline{S}$. By varying the action $\overline{S}$ with respect to $N$, $a$, and $\overline{B}_{t}$, and then setting $N=1$, $\dot{N}=0$, and $\ddot{N}=0$, we can obtain the background equations
\begin{eqnarray}
    &&\frac{1}{\kappa}\left[ 3 H^2-\Lambda+3\xi b_t^2\dot{H}+3\sigma b_t^2(3H^2+2\dot{H})-\kappa(\overline{V}+2 b_t^2 \overline{V}_{,b^2}) \right]=\overline{\rho}_{m},\label{EqN-BG}
    \\
    &&\frac{1}{\kappa}\left[ -3H^2-2\dot{H}+\Lambda+b_t^2(\xi+\sigma)(3H^2+2\dot{H})+\kappa\overline{V} \right]=\overline{p}_{m},\label{Eqa-BG}
    \\
    &&b_t\left[ -3\xi(H^2+\dot{H}) -6\sigma(2H^2+\dot{H}) +2\kappa \overline{V}_{,b^2} \right]=0,\label{EqA-BG}
\end{eqnarray}
where $\overline{V}_{,b^2}\equiv \big(\partial V/\partial b^2\big)|_{B_{\mu}=\overline{B}_{\mu}}=\big(\partial V/\partial (B^{\mu}B_{\mu}+b^2)\big)|_{B_{\mu}=\overline{B}_{\mu}}$, and we have used the definition of pressure $p_m=n\partial\rho_m/\partial n-\rho_m$. By rewriting the background equations~(\ref{EqN-BG})-(\ref{EqA-BG}), we can obtain
\begin{eqnarray}
  \xi b_{t}^2+\sigma b_t^2&=&1+\frac{\kappa}{2 \dot{H}}(\overline{p}_m+\overline{\rho}_m),\label{lorentz-xib-9242207}
  \\
  \overline{V}_{,b^2}&=&\frac{3\xi}{2\kappa}(H^2+\dot{H})+\frac{3\sigma}{\kappa}(2H^2+\dot{H}),\label{vb-9242207}
  \\
  \overline{V}&=&-\frac{\Lambda}{\kappa}+\frac{1}{\kappa}\left(1-(\xi+\sigma)b_{t}^2\right)(3H^2+2\dot{H})+\overline{p}_{m}.
\end{eqnarray}
From these equations, we can see that if the Hubble parameter $H(t)$, the matter energy density $\overline{\rho}_{m}$, and the matter pressure $\overline{p}_m$ can be determined, the sum of the Lorentz-violating parameters $\xi b^2$ and $\sigma b^2$, the vacuum potential $\overline{V}$, and $\overline{V}_{,b^2}$ can all be assigned definite values.
\par
For the Hubble parameter $H$, we consider only its non-trivial solution $H=H(t)$ in this paper. The expanding universe gives $H(t_0)>0$. The accelerating universe gives $\ddot{a}(t_0)/a(t_0)=H^2(t_0)+\dot{H}(t_0)>0$, which means $H^2(t_0)>-\dot{H}(t_0)$. We consider that $b_t$ is a non-zero constant, because the Bumblebee model is a Lorentz-violating gravity theory. Since we have not detected any Lorentz-violating effects in gravitational experiments, the Lorentz-violating parameters $\xi b^2$ and $\sigma b^2$ should be very small, $|\xi b^2|,|\sigma b^2|\ll 1$. The condition $\sigma = -\tfrac{1}{2}\xi$ is ruled out by an accelerating universe, because Eq.~\eqref{vb-9242207} would then reduce to $H^2=-2\kappa\overline{V}_{,b^2}/(3\xi)=constant$, which is observationally excluded. Therefore, we must have $\sigma \ne -\tfrac{1}{2}\xi$. Using Eq.~\eqref{lorentz-xib-9242207}, we obtain the following constraints
\begin{eqnarray}
  &&\dot{H}<0, \label{relation-H'-1611}
  \\
  &&\frac{\kappa}{2 \dot{H}}(\overline{p}_m+\overline{\rho}_m)= -1+\mathcal{O}(\xi b^2)+\mathcal{O}(\sigma b^2). \label{H'-approx-1010}
\end{eqnarray}
Here, $\mathcal{O}(\xi b^2)$ and $\mathcal{O}(\sigma b^2)$ represent all higher-order quantities than $(\xi b^2)^0$ and $(\sigma b^2)^0$, respectively. The constraint~\eqref{relation-H'-1611} is based on the premise of $\overline{\rho}_m \ne 0$. This constraint is consistent with both GR and cosmological observations~\cite{Camarena:2019moy,Feeney:2017sgx}. Since $\dot{H}$ is seldom discussed directly in cosmology, it can be derived from the deceleration parameter $q(z)$: $\dot{H}=-(1+q)H^2$. The result of Ref.~\cite{Camarena:2019moy} gives the value of the current deceleration parameter of the universe $q_{0}=-0.55$, which indicates $\dot{H}=-0.45 H^2$.

\subsection{Dark energy}
In 1998, two groups, one led by Brian Schmidt and Adam Riess~\cite{SupernovaSearchTeam:1998fmf} and the other by Saul Perlmutter~\cite{SupernovaCosmologyProject:1998vns}, obtained a surprising result: the expansion rate of the universe in the past is lower than that at the present epoch, based on the sample of 16 distant and 34 nearby supernovae. Since then, dark energy has emerged as one of the central issues in theoretical physics and cosmology. The most straightforward approach to explaining dark energy is the introduction of a cosmological constant $\Lambda$ within the framework of GR, which, from the perspective of quantum field theory, is often interpreted as vacuum energy. However, there is a significant discrepancy between the cosmological constant and the prediction by quantum field theory~\cite{Li:2011sd,Cacciatori:2024ipk}. In the Bumblebee model, if we attribute deviations from GR to dark energy, we can then discuss and analyze dark energy within this theoretical framework.
\par
We rewrite Eqs.~\eqref{EqN-BG} and~\eqref{Eqa-BG} as
\begin{eqnarray}
    \frac{3}{\kappa}H^2 &=& \overline{\rho}_m+\overline{\rho}_D,\label{H2-rho-2245}\\
    \frac{2}{\kappa}\dot{H} &=& -\overline{\rho}_m-\overline{p}_m-\overline{\rho}_D-\overline{p}_D,\label{H'-rho-2245}
\end{eqnarray}
where the specific forms of $\overline{\rho}_D$ and $\overline{p}_D$ are
\begin{eqnarray}
    \overline{\rho}_D &=& \frac{3}{\kappa}H^2-\overline{\rho}_m=\frac{\Lambda}{\kappa}+ \frac{3(\xi+\sigma) b_{t}^2}{\kappa} H^2+\overline{V}, \label{rhoD-1631}
    \\
    \overline{p}_D &=& -\frac{1}{\kappa}(3 H^2+2 \dot{H})-\overline{p}_m=-\frac{\Lambda}{\kappa}-\left[ \frac{(\xi+\sigma) b_{t}^2}{\kappa}(3 H^2+2 \dot{H})+\overline{V} \right]. \label{pD-1631}
\end{eqnarray}
Therefore, the equation of state of the dark energy is
\begin{eqnarray}
    w_D&=&\frac{\overline{p}_D}{\overline{\rho}_D}=-1-2(\xi+\sigma) b_{t}^2\frac{\dot{H}}{\kappa\overline{\rho}_D}.
\end{eqnarray}
Since $\dot{H}\ne 0$ (see relation~\eqref{relation-H'-1611}), whether $w_D$ deviates from $-1$ depends on the value of $(\xi+\sigma) b_{t}^2$. Typically, the term $-2(\xi+\sigma) b_{t}^2\dot{H}/(\kappa\overline{\rho}_D)$ always contributes to this deviation, if $b^2\ne 0$ and $(\xi+\sigma)\ne 0$. It is not difficult to see that $w_{D}$ is negative, given that $|(\xi+\sigma) b_{t}^2|\ll 1$. Thus, we have $\overline{\rho}_{D}>0$, because the right side of the first equal sign of Eq.~\eqref{pD-1631} implies $\overline{p}_{D}<0$. Combining the results $\dot{H}<0$~\eqref{relation-H'-1611} and $\xi\le 0$ (see Eq.~\eqref{Tachyon-vector-1736}), if $\sigma$ is zero, the equation of state of the dark energy satisfies
\begin{eqnarray}
  w_D\le -1.\label{wD-1191247}
\end{eqnarray}
\par
To address the dark energy problem and other related issues, many alternative approaches exist, including modified gravity theories. Modified gravity theories typically involve approaches such as introducing additional fields or incorporating higher-order derivatives as alternatives to the cosmological constant to explain dark energy. In the Bumblebee model, according to Eqs.~\eqref{rhoD-1631} and~\eqref{pD-1631}, obviously, the cosmological constant and the Bumblebee field collectively account for dark energy.
\par
According to Eqs.~\eqref{H2-rho-2245} and~\eqref{H'-rho-2245}, we obtain the equation that characterizes the accelerated expansion of the universe at present
\begin{eqnarray}
    \frac{\ddot{a}}{a}&=& H^2+\dot{H}=-\frac{\kappa}{6}\left(\overline{\rho}_m+3\overline{p}_m+\overline{\rho}_D+3\overline{p}_D\right). \label{accelerated-9111120}
\end{eqnarray}
Since $\overline{\rho}_m>0$, $\overline{p}_m>0$, and $\overline{\rho}_{D}>0$, these three terms inhibit the accelerated expansion of the universe. By analyzing Eq.~\eqref{pD-1631}, it is easy to see that the dark energy always provides a negative pressure $\overline{p}_D<0$, which drives to the accelerated expansion of the universe.

\section{The tensor perturbations}\label{tensor-9132349}
Since the equations of motion for the spatial tensor, vector, and scalar perturbations are decoupled in the cosmological background, we can analyze them separately. In this section, we focus on the tensor perturbations. Since $S=S[g_{\mu\nu},B_{\mu},J^{\mu},\ell,\mathcal{A}_1,\mathcal{A}_2,\mathcal{B}_1,\mathcal{B}_2]$, the tensor perturbations are governed solely by the metric $g_{\mu\nu}$ (see Eqs.~(\ref{ds-full-1514})-(\ref{ell-full-1514})),
\begin{eqnarray}
  ds^2=-dt^2+a^2\left(\delta_{ij}+h^{\mathrm{TT}}_{ij}\right)dx^{i}dx^{j}.
\end{eqnarray}
Here, $h^{\mathrm{TT}}_{ij}$ is a traceless and divergence-free tensor, obeying $\delta^{ij}h^{\mathrm{TT}}_{ij}=0$ and $\partial^ih^{\mathrm{TT}}_{ij}=0$. Without loss of generality, we can take the propagation direction of the GWs to be the ``$+z$" axis, so, the non-vanishing components of $h^{\mathrm{TT}}_{ij}$ are
\begin{eqnarray}
    h^{\mathrm{TT}}_{11}=-h^{\mathrm{TT}}_{22}=h_{+}(t,z),\qquad h^{\mathrm{TT}}_{12}=h^{\mathrm{TT}}_{21}=h_{\times}(t,z),
\end{eqnarray}
where $h_{+}(t,z)$ and $h_{\times}(t,z)$ characterize the two polarization states, and $|h_{+}|\ll 1,\; |h_{\times}|\ll 1$.
\par
For the Schutz-Sorkin action~\eqref{perfect-fluid-action}, the terms $J^{\mu}(\partial_{\mu}\ell+\mathcal{A}_1\partial_{\mu}\mathcal{B}_1+\mathcal{A}_2\partial_{\mu}\mathcal{B}_2)$ do not contribute to the tensor perturbations, and the expansions of $\sqrt{-g}$ and $\rho_m(n)$ are derived using perturbative methods
\begin{eqnarray}
    && \sqrt{-g}=a^3 -\frac{a^3}{2}(h_{+}^2 + h_{\times}^2) + \dots ,\\
    && \rho_m(n)=\rho_m(\overline{n}+\delta n)=\overline{\rho}_m + \frac{\overline{n}}{2}\overline{\rho}_{m,n}\left(h_{+}^2 + h_{\times}^2\right) + \dots ,
\end{eqnarray}
where $\overline{\rho}_m=\rho_m(\overline{n})$, $\overline{\rho}_{m,n}=(\partial \rho_m/\partial n)|_{n=\overline{n}}$, and ``$\dots$'' represents the higher-order terms beyond second-order perturbations. So the second-order Schutz-Sorkin action of the tensor perturbations can be written as
\begin{eqnarray}
    S^{(2)}_{m|t} =-\int d^4x\frac{a^3}{2}\overline{p}_m\left(h_{+}^2 + h_{\times}^2\right).
\end{eqnarray}
\par
By expanding the Bumblebee model action~\eqref{action-all} with a perfect fluid up to  second order in perturbations, using the background equation~\eqref{Eqa-BG}, and integrating by parts, we can finally write the total second-order action $S^{(2)}_t=S^{(2)}_{g|t}+S^{(2)}_{m|t}$ in the form
\begin{eqnarray}
    S_{t}^{(2)}=\int dtd^3x\frac{a^3}{4\kappa}\left(1-(\xi+\sigma) b_t^2\right)\left[ (\dot{h}_{+}^{2}+\dot{h}_{\times}^{2})-c_t^2\overline{g}^{zz}\big((\partial_z h_{+})^2+(\partial_z h_{\times})^2\big) \right],
\end{eqnarray}
where $c_t^2=1+(\xi  b_t^2)/(1-(\xi+\sigma) b_t^2)$ is the square of the propagation speed of tensor modes. Since $a$ and $\kappa$ are positive, to avoid Laplacian instability and ghost instability, we require
\begin{eqnarray}
  &&\frac{\xi  b_t^2}{1-(\xi+\sigma) b_t^2}>-1,\\
  &&1-(\xi+\sigma) b_t^2>0.
\end{eqnarray}
These conditions are satisfied since the Lorentz-violating parameters obey $|\xi b^2|,|\sigma b^2| \ll 1$.  Therefore, the Bumblebee model is free of both Laplacian instability and ghost instability in the tensor perturbations.
\par
Varying the action $S_t^{(2)}$ with respect to $h_+$ and $h_{\times}$, respectively, we obtain the equations of motion for the tensor perturbations
\begin{eqnarray}
    \ddot{h}_w+3H \dot{h}_w-\left(1+\frac{\xi  b_t^2}{1-(\xi+\sigma) b_t^2}\right)\overline{g}^{zz}\partial_z\partial_z h_w=0,\label{motion-hb-2340}
\end{eqnarray}
where the indicator $w$ labels $+$ and $\times$. Compared to the case in GR, there is only one deviation in the tensor perturbation equations of motion~\eqref{motion-hb-2340}: the factor $\tfrac{\xi  b_t^2}{1-(\xi+\sigma) b_t^2}$ in the third term on the left-hand side. This deviation will lead to a difference between the GW speed and the speed of light. It is straightforward to see that whether the speed of the tensor GWs deviates from the speed of light primarily depends on whether the Lorentz-violating parameter  $\xi b_t^2$ is zero. Furthermore, the stability requirement for the vector perturbations demands that $ \xi \le 0 $, resulting in the speed of tensor modes of GWs being less than 1.
\par
On August 17, 2017, a binary neutron star coalescence candidate (GW170817) was observed through GWs by Advanced LIGO and Virgo~\cite{LIGOScientific:2017vwq}. About 1.7 seconds later, the Fermi Gamma-ray Burst Monitor independently detected a gamma-ray burst (GRB170817A)~\cite{Savchenko:2017ffs}. These observations placed a tight constraint on the speed of tensor modes of GWs~\cite{LIGOScientific:2017zic, LIGOScientific:2018dkp}: $-3\times 10^{-15}\le c_t-1\le 7\times 10^{-16}$. In the Bumblebee model, the Lorentz-violating parameter can be constrained using the Taylor series expansion $c_{t}=1+\xi b_t^2/2+\mathcal{O}\big((\xi b_t^2)^n(\sigma b_t^2)^{m}\big)$, where $n+m\ge 2$,
\begin{eqnarray}
  -6\times 10^{-15} \lesssim \xi b_t^2 \lesssim 1.4\times 10^{-15}.
\end{eqnarray}
Here, the ``$\lesssim$" symbol arises from the small quantity $\mathcal{O}\big((\xi b_t^2)^n(\sigma b_t^2)^{m}\big)$. Furthermore, the stability requirement for the vector perturbations $\xi\le 0$ given in Eq.~\eqref{Tachyon-vector-1736} places an additional constraint on the Lorentz-violating parameter
\begin{eqnarray}
  -6\times 10^{-15} \lesssim \xi b_t^2 \lesssim 0.\label{xibt-9112245}
\end{eqnarray}
This inequality indicates that the Lorentz-violating parameter $\xi b_t^2$ is negative and is constrained at the $10^{-15}$ level.
\par
There are many other studies constraining the Lorentz-violating parameter in the Bumblebee model. Reference~\cite{Casana:2017jkc} constrains the upper bound of the Lorentz-violating parameter at the $10^{-13}$ level by time delay of light. Reference~\cite{Wang:2021gtd} indicates that the Lorentz-violating parameter should be negative by studying quasi-periodic oscillations frequencies in a rotating black hole. Both references consider the vacuum expectation value of the Bumblebee field to be $(0,b(r),0,0)$. For the case of a time-like vacuum expectation value $(B(t),0,0,0)$, Ref.~\cite{Khodadi:2022mzt} constrains the Lorentz-violating parameter by analyzing the data from Big Bang Nucleosynthesis and Gravitational Baryogenesis. This result shows that the strength of the constraint on the Lorentz-violating parameter depends on the choice of the exponent parameter $\beta$. If $-0.038< \beta <0$, the Lorentz-violating parameter is positive and can be constrained at the $10^{-24}$ level. Our work differs from Ref.~\cite{Khodadi:2022mzt}, which employs the ansatz $B(t)\sim t^{\beta}$ for the vector field's time evolution. For further related studies, see Refs.~\cite{Bertolami:2005bh,Paramos:2014mda}.

\section{The vector perturbations} \label{vector-section-9132359}
\subsection{The second-order action of the vector perturbations}
In this section, we focus on the vector perturbations. The tensor and scalar perturbations are not considered here.
According to the scalar-vector-tensor decomposition, the fields $g_{\mu\nu}$, $B_{\mu}$, $J^{\mu}$, $\mathcal{A}_1$, $\mathcal{A}_2$, $\mathcal{B}_1$, and $\mathcal{B}_2$ give rise to the vector perturbations. The specific forms of the perturbed line element, vector field, and vector density are (see Eqs.~(\ref{ds-full-1514})-(\ref{ell-full-1514}))
\begin{eqnarray}
    && ds^2=-dt^2+2\lambda_i dx^idt +a^2\left[\delta_{ij}+(\partial_i\varepsilon_j +\partial_j\varepsilon_i)\right]dx^idx^j, \label{Vds-1726}\\
    && B_{\mu}=\left(b_{t},\; \varsigma_i\right),\\
    && J^{\mu}=\left(\overline{J},\; \chi^i\right).\label{VJ-1727}
\end{eqnarray}
Here, the perturbations $\lambda_i$, $\varepsilon_i$, $\varsigma_i$, and $\chi^{i}$ are functions of the space-time coordinates, and $\partial^{i}\lambda_i=\partial^{i}\varepsilon_i=\partial^{i}\varsigma_i=\partial_{i}\chi^i=0$. Without loss of generality, we can always choose the propagation direction of perturbations as the ``$+z$" direction. Thus,  $\lambda_i=\lambda_i(t,z)$, $\varepsilon_i=\varepsilon_i(t,z)$, $\varsigma_i=\varsigma_i(t,z)$, and $\chi^{i}=\chi^{i}(t,z)$, where $\lambda_z=\varepsilon_z=\varsigma_z=\chi^z=0$. The specific forms of perturbations for $\mathcal{A}$ and $\mathcal{B}$ are given in Eq.~\eqref{AB-Perturbations-1117}.
\par
Expanding the Schutz-Sorkin action~\eqref{perfect-fluid-action} up to second order in the vector perturbations, we obtain the second-order perturbation action for the perfect fluid as follows
\begin{eqnarray}
  S_{m|v}^{(2)}=\int dt d^3x \left[ -\frac{a^3}{2}\overline{p}_{m}\delta^{pq}\partial_z\varepsilon_{p}\partial_z\varepsilon_{q} +\frac{a}{2}\overline{p}_{m}\delta^{pq}\lambda_{p}\lambda_{q} +\frac{a^2}{2 \overline{J}}\overline{\rho}_{m,n}\delta_{pq}\chi^{p}\chi^{q} +\overline{\rho}_{m,n}\lambda_{p}\chi^{p} -(\chi^{p}+\delta^{pq}\overline{J}\dot{\delta\mathcal{B}}_{q})\delta\mathcal{A}_{p} \right].\label{Smv20-1555}
\end{eqnarray}
Here, the indices $p$ and $q$ run over 1 and 2. Varying $S_{m|v}^{(2)}$ with respect to $\chi^{p}$, $\mathcal{A}_{p}$, and $\mathcal{B}_{p}$, respectively, we can obtain the perturbation equations of matter
\begin{eqnarray}
  && \delta\mathcal{A}_{p} -\overline{\rho}_{m,n}\lambda_{p} -\frac{a^2 \overline{\rho}_{m,n}}{\overline{J}}\delta_{pq}\chi^{q}=0, \label{chi-eq1-1546}
  \\
  && \delta_{pq}\chi^{q} +\overline{J}\dot{\delta\mathcal{B}}_{p}=0, \label{A-eq1-1546}
  \\
  && \overline{J}\dot{\delta\mathcal{A}}_{p}=0. \label{B-eq1-1546}
\end{eqnarray}
Equation~\eqref{B-eq1-1546} implies that $\delta\mathcal{A}_{p}$ is a function of $z$ only, i.e., $\delta\mathcal{A}_{p}=\delta\mathcal{A}_{p}(z)$. Substituting the constraint~\eqref{chi-eq1-1546} into the action~\eqref{Smv20-1555} and eliminating the variable $\delta\mathcal{A}_{p}$, we obtain an effective action. Furthermore, by varying this effective action with respect to $\chi^{p}$, we can obtain a constraint on $\chi^{p}$, 
\begin{eqnarray}
  \delta_{pq}\chi^{q} +\overline{J}\dot{\delta\mathcal{B}}_{p}=0. \label{chi-eq2-1806}
\end{eqnarray}
By substituting this constraint into the effective action, the variable $\chi^{p}$ is eliminated. Finally, the second-order matter action reduces to
\begin{eqnarray}
  S_{m|v}^{(2)}=\int dtd^3x \left[ \frac{\overline{J}a^2}{2 }\overline{\rho}_{m,n}\delta^{pq}\dot{\delta\mathcal{B}}_{p}\dot{\delta\mathcal{B}}_{q} +\frac{a}{2}\overline{p}_{m}\delta^{pq}\lambda_{p}\lambda_{q} -\frac{a^3}{2}\overline{p}_{m}\delta^{pq}\partial_{z}\varepsilon_{p}\partial_{z}\varepsilon_{q} -\overline{J}\overline{\rho}_{m,n}\delta^{pq}\dot{\delta\mathcal{B}}_{p}\lambda_{q} \right].
\end{eqnarray}
Combining this second-order matter action and expanding the action~\eqref{action-all} up to second order in perturbations, we can obtain the full second-order perturbation action
\begin{eqnarray}
  S^{(2)}_{v}=\int dtd^3x &\Bigg[& \frac{\overline{J}a^2}{2 }\overline{\rho}_{m,n}\delta^{pq}\dot{\delta\mathcal{B}}_{p}\dot{\delta\mathcal{B}}_{q} +\frac{a}{2}\delta^{pq}\dot{\varsigma}_{p}\dot{\varsigma}_{q} -\frac{1}{2 a}\delta^{pq}\partial_{z}\varsigma_{p}\partial_{z}\varsigma_{q} -\frac{\xi a}{\kappa}H'\delta^{pq}\varsigma_{p}\varsigma_{q} \nonumber
  \\
  && +\frac{1-(\xi+\sigma) b_{t}^2}{4\kappa a}\delta^{pq}\partial_{z}\lambda_{p}\partial_{z}\lambda_{q} +\frac{\overline{J}}{2 a^2}\overline{\rho}_{m,n}\delta^{pq}\lambda_{p}\lambda_{q} -\overline{J}\overline{\rho}_{m,n}\delta^{pq}\dot{\delta\mathcal{B}}_{p}\lambda_{q} -\frac{\xi b_{t}}{2\kappa a}\delta^{pq}\partial_{z}\lambda_{p}\partial_{z}\varsigma_{q} +\mathcal{L}_{\varepsilon} \;\;\Bigg], \label{S2-action-1759}
\end{eqnarray}
where
\begin{eqnarray}
  \mathcal{L}_{\varepsilon}=\frac{1-(\xi+\sigma) b_{t}^2}{4\kappa}a^3 \delta^{pq}\partial_{z}\dot{\varepsilon}_{p}\partial_{z}\dot{\varepsilon}_{q} +\frac{a}{2\kappa}\delta^{pq}\partial^2_{z}\left[ \xi b_{t}(H \varsigma_{p}+\dot{\varsigma}_{p})-\left(1-(\xi+\sigma) b_{t}^2\right)\big(H\lambda_{p}+\dot{\lambda}_{p}\big) \right]\varepsilon_{q}.
\end{eqnarray}
Here, we have used the background equations~(\ref{EqN-BG})-(\ref{EqA-BG}) and integration by parts. Note that the action~\eqref{S2-action-1759} is not the original second-order perturbation action, since the variables $\delta\mathcal{A}_{p}$ and $\chi^{p}$ have been eliminated. However, if the Lagrange multiplier terms for the constraints \eqref{chi-eq1-1546} and \eqref{chi-eq2-1806} are considered, the action~\eqref{S2-action-1759} is equivalent to the original one. Varying the action~\eqref{S2-action-1759} with respect to $\lambda_{p}$, $\varepsilon_{p}$, $\varsigma_{p}$, and $\delta\mathcal{B}_{p}$, respectively, we obtain the perturbation equations
\begin{eqnarray}
  && -\overline{J}\overline{\rho}_{m,n}\dot{\delta\mathcal{B}}_{p} +\frac{1-(\xi+\sigma) b_{t}^2}{2\kappa}a\partial_z^2\dot{\varepsilon}_{p} +\frac{\overline{J}}{a^2}\overline{\rho}_{m,n}\lambda_{p} -\frac{1-(\xi+\sigma) b_{t}^2}{2\kappa a}\partial_z^2\lambda_{p} +\frac{\xi b_{t}}{2\kappa a}\partial_z^2\varsigma_{p}=0,\label{motion1-vector-2117}
  \\
  && \partial_z^2\left( a^2\big(1-(\xi+\sigma) b_{t}^2\big)(\ddot{\varepsilon}_{p}+3H\dot{\varepsilon}_{p}) +\xi b_{t}(\dot{\varsigma}_{p}+H\varsigma_{p})-\big(1-(\xi+\sigma) b_{t}^2\big)(\dot{\lambda}_{p}+H\lambda_{p}) \right)=0,
  \\
  && \ddot{\varsigma}_{p} +H\dot{\varsigma}_{p} +\frac{2\xi}{\kappa}\dot{H}\varsigma_{p} -\frac{1}{a^2}\partial_z^2\varsigma_{p} -\frac{\xi b_{t}}{2\kappa a^2}\partial_z^2\lambda_{p} +\frac{\xi b_{t}}{2\kappa}\partial_z^2\dot{\varepsilon}_{p}=0,
  \\
  && a^2\overline{\rho}_{m,n}\ddot{\delta\mathcal{B}}_{p} +a^2 H(2\overline{\rho}_{m,n}-3\overline{p}_{m,n})\dot{\delta\mathcal{B}}_{p} -\overline{\rho}_{m,n}\dot{\lambda}_{p} +3 H\overline{p}_{m,n}\lambda_{p}=0.\label{motion4-vector-2117}
\end{eqnarray}

\subsection{Gauge issues, effective action, and stability conditions}\label{gaugeSec-vector-2109}
For analyzing the dynamical behavior of the Bumblebee system, we must eliminate all gauge degrees of freedom. To facilitate the analysis, we separate the non-dynamical variables from the action. In this subsection, we perform these two operations to obtain an effective action, and finally perform the stability analysis of the Bumblebee model.
\par
Since the Bumblebee model is a covariant theory, the linearized theory is invariant under infinitesimal local coordinate transformations. Let us consider a infinitesimal local coordinate transformation that acts on the vectors,
\begin{eqnarray}
    x^{\mu}\rightarrow x^{\mu}+\xi^{\mu},\qquad \xi^{\mu}=(0,\; \xi_{\mathrm{T}}^i),\label{gaugeTransformation-x-1759}
\end{eqnarray}
where $|\xi_{\mathrm{T}}^i|\ll 1$, $\partial_i \xi_{\mathrm{T}}^i =0$, and $\xi_{\mathrm{T}}^i$ is a function of the space-time coordinates $x^\mu$. Under this transformation, the perturbations of the metric, vector field, and vector density transform as follows
\begin{eqnarray}
    \lambda_i &\rightarrow& \lambda_i -a^2\delta_{ik}\dot{\xi}_{\mathrm{T}}^{k},\label{transformations-perturbation1-2124}\\
    \varepsilon_i &\rightarrow& \varepsilon_i -\delta_{ik}\xi_{\mathrm{T}}^k,\\
    \varsigma_i &\rightarrow& \varsigma_i,\\
    \chi^{i} &\rightarrow& \chi^{i}+\overline{J}\dot{\xi}_{\mathrm{T}}^i.\label{transformations-perturbation4-2124}
\end{eqnarray}
Since the linearized theory is gauge-invariant, we can always fix the values of certain quantities among $\lambda_i$, $\varepsilon_i$, $\varsigma_i$, and $\chi^i$ based on the perturbation transformations~\eqref{transformations-perturbation1-2124}-\eqref{transformations-perturbation4-2124} without altering the physical outcome. If we select the gauge condition $\lambda_i=0$ or $\chi^{i}=0$, the transformation vector $\xi_{\mathrm{T}}^i$ is not unambiguously fixed. Since $\dot{\xi}_{\mathrm{T}}^i=\dot{\xi}_{f}^i$, where $\xi_{f}^i=\xi_{\mathrm{T}}^i+f_{\mathrm{T}}^i(x,y,z)$, there is still gauge invariance under the vector transformation $f_T^i(x,y,z)$, indicating that the gauge freedom is not fully fixed. Hence, we select the gauge condition
\begin{eqnarray}
  \varepsilon_i=0.\label{gauge-vector-1635}
\end{eqnarray}
Then we will analyze the stability of the linearized theory under the vector perturbations with the gauge condition $\varepsilon_i=0$.
\par
In the action~\eqref{S2-action-1759}, it is easy to see that there is no kinetic term for the variable $\lambda_{p}$, which implies that it provides a constraint equation. In Fourier space, substituting the gauge condition $\varepsilon_i=0$ into the action~\eqref{S2-action-1759} and varying it with respect to $\lambda_{p}$, we can obtain a constraint equation
\begin{eqnarray}
  -\overline{J}\overline{\rho}_{m,n}\dot{\delta\mathcal{B}}_{p} +\left(\frac{\overline{J}}{a^2}\overline{\rho}_{m,n} +\frac{1-(\xi+\sigma) b_{t}^2}{2\kappa a}k_z^2 \right)\lambda_{p} -\frac{\xi b_{t}}{2\kappa a}k_z^2\varsigma_{p}=0,\label{lambda-1622}
\end{eqnarray}
where $k_z$ is a wavenumber. Substituting this constraint into the action~\eqref{S2-action-1759} in Fourier space, we can eliminate the non-dynamical variable $\lambda_{p}$ and obtain an effective action in Fourier space
\begin{eqnarray}
  S^{(2)}_{v}=\int dtd^3x &\Bigg[& \frac{a}{2}\delta^{pq}\dot{\varsigma}_{p}\dot{\varsigma}_{q} -\left( \frac{1}{2 a}+\frac{\xi^2 b_{t}^2 k_z^2}{4\kappa\big(1-(\xi+\sigma) b_{t}^2\big)a k_z^2+8\kappa^2\overline{J}\overline{\rho}_{m,n}} \right)k_z^2 \delta^{pq}\varsigma_{p}\varsigma_{q} -\frac{\xi a}{\kappa}\dot{H}\delta^{pq}\varsigma_{p}\varsigma_{q}  \nonumber
  \\
  && +\frac{\overline{J}\big(1-(\xi+\sigma) b_{t}^2\big)a^3\overline{\rho}_{m,n}k_z^2}{2\big(1-(\xi+\sigma) b_{t}^2\big)a k_z^2+4\kappa\overline{J}\overline{\rho}_{m,n}}\delta^{pq}\dot{\delta\mathcal{B}}_{p}\dot{\delta\mathcal{B}}_{q} -\frac{\xi b_{t}\overline{J}a\overline{\rho}_{m,n}k_{z}^2}{\big(1-(\xi+\sigma\big) b_{t}^2)a k_{z}^2+2\kappa\overline{J}\overline{\rho}_{m,n}}\delta^{pq}\varsigma_{p}\dot{\delta\mathcal{B}}_{q} \;\;\Bigg].\label{effectiveAction1-1314}
\end{eqnarray}
It is not difficult to see that the variables $\delta\mathcal{B}_{1}$ and $\delta\mathcal{B}_{2}$ are cyclic coordinates. They correspond to two conserved quantities
\begin{eqnarray}
  \frac{\overline{J}a\overline{\rho}_{m,n}k_z^2}{\big(1-(\xi+\sigma) b_{t}^2\big)a k_z^2+2\kappa\overline{J}\overline{\rho}_{m,n}} \left( \xi b_{t}\varsigma_{p} -\big(1-(\xi+\sigma) b_{t}^2\big)a^2\dot{\delta\mathcal{B}}_{p} \right) =C^{v}_{p}, \label{conserved-1312}
\end{eqnarray}
where $C^{v}_{p}$ are constants. Since we are only concerned with the dynamic effects, we can choose $C_{p}^{v}=0$ without altering the physics. For a more general case, we can solve for $C_p^{v}$, which depends non-locally on $\varsigma_{p}$ (see Ref.~\cite{Dyer:2008hb}). However, we are not concerned with this general case. Substituting this equation~\eqref{conserved-1312} into the action~\eqref{effectiveAction1-1314} and eliminating $\dot{\delta\mathcal{B}}_{p}$, we obtain an effective action containing only dynamical variables
\begin{eqnarray}
  S^{(2)}_{v}=\int dtd^3x \frac{a}{2}\left[ \delta^{pq}\dot{\varsigma}_{p}\dot{\varsigma}_{q} -c^2_{v}\overline{g}^{zz}k_z^2\delta^{pq}\varsigma_{p}\varsigma_{q} -m^2_{v}\delta^{pq}\varsigma_{p}\varsigma_{q} \right].\label{effectiveAction2-vector-1742}
\end{eqnarray}
where $c^2_{v}=1+ \xi^2 b_{t}^2/\big( 2\kappa \big(1-(\xi+\sigma) b_{t}^2\big) \big)$ and $m^2_{v}=2\xi \dot{H}/\kappa$. $c^2_{v}$ and $m^2_{v}$ are the vector propagation speed squared and effective mass squared, respectively. It is easy to see that there are two dynamical degrees of freedom $\varsigma_{1}$ and $\varsigma_{2}$ for the vector perturbations. In Fourier space, this action is equivalent with the original one under the gauge condition~\eqref{gauge-vector-1635} when the constraints~\eqref{chi-eq1-1546}, \eqref{chi-eq2-1806}, \eqref{lambda-1622}, and \eqref{conserved-1312} are considered (with $C_p^{v}=0$). When the boundary conditions are specified, these constraints imply that the non-dynamical variables $\lambda_p$, $\varepsilon_p$, $\chi^{p}$, $\delta\mathcal{A}_p$, and $\delta\mathcal{B}_p$ all depend on the variable $\varsigma_{p}$.
\par
The action~\eqref{effectiveAction2-vector-1742} is an effective action that only contains the dynamical variable $\varsigma_{p}$. From this action, the stability conditions for the vector perturbations in the Bumblebee model can be readily derived. First, since $a>0$, the vector perturbations are ghost-free. Second, the conditions for avoiding Laplacian instability and tachyonic instability are as follows
\begin{eqnarray}
  && c^2_{v}=1+ \frac{\xi^2 b_{t}^2}{2\kappa \big(1-(\xi+\sigma) b_{t}^2\big)} >0, \\
  && m^2_{v}=\frac{2\xi}{\kappa}\dot{H} \ge 0.
\end{eqnarray}
The condition $|\xi b_{t}^2|,|\sigma b_{t}^2|\ll1$ for the Lorentz-violating parameters ensure Laplacian stability. To avoid tachyonic instability, and considering that $\dot{H}<0$ (see Eq.~\eqref{relation-H'-1611}), the following requirement must be met
\begin{eqnarray}
  \xi\le 0.\label{Tachyon-vector-1736}
\end{eqnarray}
This condition further tightens the bound on the speed of the tensor GW in the Bumblebee model, see Eq.~\eqref{xibt-9112245}.
\par
Varying the action~\eqref{effectiveAction2-vector-1742} with respect to $\varsigma_{q}$, we can obtain the equation of motion for $\varsigma_{q}$
\begin{eqnarray}
  \ddot{\varsigma}_{p} +H\dot{\varsigma}_{p} +c^2_{v}\overline{g}^{zz}k_z^2\varsigma_{p} +m^2_{v}\varsigma_{p}=0.\label{motion-vector-9261532}
\end{eqnarray}
In the small-scale limit ($k_z \rightarrow \infty$), it is easy to see that the dispersion relation is 
\begin{eqnarray}
  w_{v}^2-c^2_{v}\overline{g}^{zz}k_z^2=0,
\end{eqnarray}
where $w_{v}$ is the frequency of the vector perturbations. Since $\kappa>0$ and $|\xi b_{t}^2|,|\sigma b_{t}^2|\ll1$, the propagation speed of the vector perturbations is greater than $1$, which is caused by the violation of Lorentz symmetry. If the Lorentz-violating parameter $\xi b_{t}^2$ equals zero, the speed will return to $1$.
\par
In this section, we analyze the dynamics of the vector perturbations in the Bumblebee model under the gauge condition $\varepsilon_i=0$~\eqref{gauge-vector-1635}. For the vector perturbations, there are two dynamical degrees of freedom $\varsigma_{1}$ and $\varsigma_{2}$~\eqref{effectiveAction2-vector-1742}. The propagation speed of the vector perturbations is greater than $1$ due to the violation of Lorentz symmetry. Regarding stability of the vector perturbations, the conditions for Laplacian stability and the free of ghost are easy to be satisfied. However, avoiding tachyonic instability requires $\xi\le 0$.

\section{The scalar perturbations} \label{scalar-section-91409}
\subsection{The second-order action of the scalar perturbations}
So far, we have analyzed the tensor and vector perturbations of the Bumblebee model, respectively. We now focus on the scalar perturbations, analyzing the dynamics and the constraints on the parameters imposed by the stability conditions.
\par
The full action~\eqref{action-all} is a functional of $g_{\mu\nu}$, $B_{\mu}$, $J^{\mu}$, $\ell$, $\mathcal{A}_1$, $\mathcal{A}_2$, $\mathcal{B}_1$, and $\mathcal{B}_2$. Since $\mathcal{A}_1$, $\mathcal{A}_2$, $\mathcal{B}_1$, and $\mathcal{B}_2$ contribute only to the vector perturbations of matter, it is easy to see that $g_{\mu\nu}$, $B_{\mu}$, $J^{\mu}$, and $\ell$ give rise to the scalar perturbations. The specific forms of the perturbations are (see Eqs.~(\ref{ds-full-1514})-(\ref{ell-full-1514}))
\begin{eqnarray}
  &&ds^2=-(1+2\phi_{h})dt^2+2\partial_{i}\varphi_{h}dx^{i}dt+a^2\left[ \delta_{ij}+E\delta_{ij}+\partial_{i}\partial_{j}\alpha \right]dx^{i}dx^{j},
  \\
  &&B_{\mu}=\overline{B}_{\mu}+(\phi_{b},\; \partial_{i}\varphi_{b}),
  \\
  &&J^{\mu}=\overline{J}^{\mu}+\left(\phi_{m},\; \frac{1}{a^2}\delta^{ij}\partial_{j}\varphi_{m}\right), 
  \\
  &&\ell=\overline{\ell}+\phi_{\ell}. 
\end{eqnarray}
Here, there are nine scalar perturbations $(\phi_{h}, \varphi_{h}, E, \alpha, \phi_{b}, \varphi_{b}, \phi_{m}, \varphi_{m}, \phi_{\ell})$, which are functions of the space-time coordinates. Substituting these perturbations into the full action~\eqref{action-all}, applying integration by parts, and using the background equations~(\ref{EqN-BG})-(\ref{EqA-BG}), we can derive the second-order perturbation action
\begin{eqnarray}\label{action-second-1030}
  S^{(2)}_{s}=\int dtd^3x &\Bigg[& -\frac{a^3}{2\kappa}\left(3\big(2-(\xi-8\sigma) b_{t}^2\big)H^2+9(\xi+2\sigma) b_{t}^2 \dot{H}+4\kappa b_{t}^4 \overline{V}_{,b^2b^2}\right)\phi_{h}^2 +\frac{(\xi+2\sigma) b_{t}^2}{\kappa}a(\partial \phi_{h})^2 
    \nonumber\\
  && +\left( \frac{b_{t}}{\kappa}a^3\big(-3(\xi-2\sigma) H^2+3(\xi+2\sigma) \dot{H}+4\kappa b_{t}^2\overline{V}_{,b^2b^2}\big)\phi_{b} +\frac{(\xi+2\sigma) b_{t}}{\kappa}a\partial^2\phi_{b} -\frac{3(\xi+2\sigma) b_{t}}{\kappa}a^3H\dot{\phi}_{b} \right)\phi_{h}
    \nonumber\\
  && -2b_{t}^2\overline{V}_{,b^2b^2}a^3\phi_{b}^2 +\frac{a}{2}(\partial \phi_{b})^2 -\frac{\overline{\rho}_{m,nn}}{2 a^3}\phi_{m}^2 +\frac{\overline{\rho}_{m,n}}{2\overline{J}a^2}(\partial \varphi_{m})^2 -(\overline{\rho}_{m,n}\phi_{h}+\dot{\phi}_{\ell})\phi_{m} +\frac{1}{a^2}\varphi_{m}\partial^2\phi_{\ell}
    \nonumber\\
  && +\mathcal{L}_{\alpha} +\mathcal{L}_{E} +\mathcal{L}_{\varphi_{b}} +\mathcal{L}_{\varphi_{h}} \;\;\Bigg].
\end{eqnarray}
This is a gauge ready form of the second-order perturbation action according to the gauge conditions~\eqref{gauge-3kind-2057}. The specific forms of the Lagrangian components $\mathcal{L}_{\alpha}$, $\mathcal{L}_{E}$, $\mathcal{L}_{\varphi_{b}}$, and $\mathcal{L}_{\varphi_{h}}$ are
\begin{eqnarray}
  \mathcal{L}_{\alpha}&=& -\frac{\overline{J}^2\overline{\rho}_{m,nn}}{8 a^3}(\partial^2\alpha)^2 +\frac{(\xi+2\sigma) b_{t}}{2\kappa}a^3\big(b_{t}\partial^2\phi_{h}-\partial^2\phi_{b}\big)\ddot{\alpha} +\frac{a^3}{\kappa}H\big((1+3\sigma b_{t}^2)\partial^2\phi_{h}-(\xi+4\sigma) b_{t}\partial^2\phi_{b}\big)\dot{\alpha} \nonumber\\
   && -\left( \frac{1-(\xi+\sigma) b_{t}^2}{\kappa}a^3\dot{H}\partial^2\phi_{h} +\frac{3\overline{J}^2\overline{\rho}_{m,nn}}{4 a^3}\partial^2E -\frac{1-(\xi+\sigma) b_{t}^2}{2\kappa}\partial^2\partial_{t}(a^3\dot{E})  -\frac{\overline{J}\overline{\rho}_{m,nn}}{2 a^3}\partial^2\phi_{m} \right)\alpha,
  \\
  \mathcal{L}_{E}&=& \frac{(1-\sigma b_{t}^2)a}{4\kappa}(\partial E)^2 -\frac{3\big(1-(\xi+\sigma) b_{t}^2\big)}{4\kappa}a^3\dot{E}^2 -\frac{9\overline{J}^2\overline{\rho}_{m,nn}}{8 a^3}E^2 +\frac{3(\xi+2\sigma) b_{t}}{2\kappa}a^3(b_{t}\phi_{h}-\phi_{b})\ddot{E} \nonumber\\
    && -\left( \frac{3\big(1-(\xi+\sigma) b_{t}^2\big)}{\kappa}a^3\dot{H}\phi_{h}+\frac{(1+\sigma b_{t}^2)a}{\kappa}\partial^2\phi_{h} -\frac{2\sigma b_{t} a}{\kappa}\partial^2\phi_{b} -\frac{3\overline{J}\overline{\rho}_{m,nn}}{2 a^3}\phi_{m} \right)E  \nonumber\\
    && +\frac{1}{\kappa}\left( 3a^3H\big((1+3\sigma b_{t}^2)\phi_{h}-(\xi+4\sigma) b_{t}\phi_{b}\big)+\big(1-(\xi+\sigma) b_{t}^2\big)a\partial^2\varphi_{h}-\xi b_{t}a\partial^2\varphi_{b} \right)\dot{E},
  \\
  \mathcal{L}_{\varphi_{b}}&=& \frac{1}{2}a(\partial \dot{\varphi}_{b})^2 -\frac{\xi}{\kappa}a \dot{H}(\partial \varphi_{b})^2 +a(\partial^2\phi_{b}) \dot{\varphi}_{b} +\frac{2\xi b_{t}}{\kappa}a H(\partial^2\phi_{h})\varphi_{b},
  \\
  \mathcal{L}_{\varphi_{h}}&=& -\frac{1-(\xi+\sigma) b_{t}^2}{\kappa}a \dot{H}(\partial \varphi_{h})^2 -\left( \frac{2(1-(\xi-\sigma) b_{t}^2)}{\kappa}a H\partial^2\phi_{h} -\frac{4\sigma b_{t}}{\kappa}a H\partial^2\phi_{b} +\frac{\overline{\rho}_{m,n}}{a^2}\partial^2\varphi_{m} \right)\varphi_{h}\nonumber\\
    && -\frac{(\xi+2\sigma) b_{t}}{\kappa}a(b_{t}\partial^2\phi_{h}-\partial^2\phi_{b})\dot{\varphi}_{h}.
\end{eqnarray}
Here, $\partial^2=\delta^{ij}\partial_{i}\partial_{j}$ and $(\partial f)^2=\delta^{ij}\partial_{i}f\partial_{j}f$, where $f$ represents any variable. Varying this action with respect to the perturbations $(\phi_{h}, \varphi_{h}, E, \alpha, \phi_{b}, \varphi_{b}, \phi_{m}, \varphi_{m}, \phi_{\ell})$, respectively, their equations of motion are
\begin{eqnarray}
  Q_{\phi_{h}}=0,\quad Q_{\varphi_{h}}=0,\quad Q_{E}=0,\quad Q_{\alpha}=0,\quad Q_{\phi_{b}}=0,\quad Q_{\varphi_{b}}=0,\quad Q_{\phi_{m}}=0,\quad Q_{\varphi_{m}}=0,\quad Q_{\phi_{\ell}}=0.\label{equation-scalar-phil-2147}
\end{eqnarray}
The specific forms of $Q_{\bullet}$ are presented in appendix \ref{appendix-1-9122141}.
\par
The covariance of the Bumblebee model implies that the linearized theory is invariant under infinitesimal local coordinate transformations. To analyze the dynamical behavior of the Bumblebee model, we must eliminate all gauge freedoms. Let us consider a infinitesimal local coordinate transformation which contributes to the scalars,
\begin{eqnarray}
  x^{\mu} \rightarrow x^{\mu}+\xi^{\mu},\qquad \xi^{\mu}=(\xi^t,\;\delta^{ij}\partial_{j}C), \label{transformation-scalar-2001}
\end{eqnarray}
where $|\xi^t|\ll 1$ and $|C|\ll 1$. The quantities $\xi^t$ and $C$ are arbitrary functions of the space-time coordinates $x^\mu$. Under this transformation, the transformations of the perturbations $(\phi_{h}, \varphi_{h}, E, \alpha, \phi_{b}, \varphi_{b}, \phi_{m}, \varphi_{m}, \phi_{\ell})$ are as follows
\begin{subequations}\label{transformation-scalars-2047}
\begin{align}
    & \phi_{h}\rightarrow \phi_{h}-\dot{\xi}^{t}, \quad \varphi_{h}\rightarrow \varphi_{h}+\xi^{t}-a^2\dot{C}, \quad E\rightarrow E-2\dot{H}\xi^{t}, \quad \alpha\rightarrow \alpha-2C  ,
    \\
    & \phi_{b}\rightarrow \phi_{b}-b_{t}\dot{\xi}^t, \quad \varphi_{b}\rightarrow \varphi_{b}-b_{t}\xi^t ,
    \\
    & \phi_{m}\rightarrow \phi_{m}-\overline{J}\partial^2C, \quad \varphi_{m}\rightarrow \varphi_{m}+a^2\overline{J}\dot{C} .
\end{align}
\end{subequations}
According to the transformation~\eqref{transformation-scalar-2001}, it is easy to see that there are two gauge degrees of freedom for the scalar perturbations in the Bumblebee model. Since $\xi^t$ and $C$ are arbitrary functions of the space-time coordinates, we can always choose suitable values for them, such that the values of certain perturbation quantities are fixed after the transformations~\eqref{transformation-scalars-2047}. A convenient gauge choice is to set the fixed scalars to zero. This does not alter the physical outcome. Similar to Sec.~\ref{gaugeSec-vector-2109}, to fix all the gauge degrees of freedom, we consider the following three kinds of gauge conditions
\begin{subequations}\label{gauge-3kind-2057}
\begin{align}
  &\textbf{Gauge I:}\quad \alpha=0, E=0.\\
  &\textbf{Gauge II:}\quad \alpha=0, \varphi_{h}=0.\\
  &\textbf{Gauge III:}\quad \alpha=0, \varphi_{b}=0.
\end{align}
\end{subequations}
Next, we will derive the stability conditions and constrain the parameter space of the Bumblebee model by applying the three gauge conditions. We will then demonstrate that different choices of the gauge conditions lead to the same physical conclusions.

\subsection{Effective action and stability conditions} \label{Gauge1-scalar-9252232}
We begin with \textbf{the gauge condition} \bm{$\alpha=0,\; E=0$}. In Fourier space, we focus on obtaining an effective action that contains only the dynamical variables, deriving the stability conditions, and determining the constraints on the parameter space.
\par
Varying the action~\eqref{action-second-1030} with respect to $\varphi_{m}$, $\phi_{\ell}$, and $\varphi_{h}$, respectively, we can obtain the constraints for them in Fourier space
\begin{eqnarray}
  &&\phi_{\ell}=\overline{\rho}_{m,n} \Big( \varphi_{h}+\frac{1}{\overline{J}}\varphi_{m} \Big),\label{phiell-912}\\
  &&\varphi_{m}=\frac{a^2}{k^2}\dot{\phi}_{m},\\
  && \varphi_{h}=\Big( (2-3\xi b_{t}^2)H\phi_{h} -(\xi+2\sigma)b_{t}^2\dot{\phi}_{h} +(\xi-2\sigma)b_{t}H\phi_{b} +(\xi+2\sigma)b_{t}\dot{\phi}_{b} +\frac{\kappa \overline{\rho}_{m,n}}{a^3}\varphi_{m} \Big)/\Big( 2\dot{H}\big(1-(\xi+\sigma) b_{t}^2\big) \Big),\label{phih-912}
\end{eqnarray}
where $k$ is a wavenumber, $k^2=\delta^{ij}k_{i}k_{j}$. By substituting these three constraints one by one into the action~\eqref{action-second-1030} in Fourier space, we can eliminate the variables $\varphi_{m}$, $\phi_{\ell}$, and $\varphi_{h}$. In order to eliminate the coupling terms between the time derivatives of the variables, which greatly facilitates the identification of non-dynamic variables, we define a new variable
\begin{eqnarray}
  \psi_1\equiv \phi_{b} -b_{t}\phi_{h} +\frac{\kappa \overline{\rho}_{m,n}}{(\xi+2\sigma)b_{t}a k^2}\phi_{m}.\label{psi1-912}
\end{eqnarray}
The variable $\phi_{b}$ can then be eliminated by $\psi_1$, yielding an effective action. This action is equivalent to the original one. Since it serves merely as an auxiliary action, we will not present its specific form here. Likewise, any subsequent auxiliary actions will also be omitted.
\par
By analyzing the newly obtained action, one can determine whether any non-dynamical perturbations remain by checking if all perturbations possess kinetic terms. In our newly obtained action, the perturbation $\phi_{h}$ leads to a constraint
\begin{eqnarray}
  \phi_{h}&=&\Big( \kappa b_{t}\dot{H}k^2\dot{\varphi}_{b} +2\xi b_{t}H\dot{H}k^2\varphi_{b} -(\xi+2\sigma)(k^2-3 a^2\dot{H})b_{t}H\dot{\psi}_{1} -\frac{3\kappa\overline{\rho}_{m,n}}{k^2}a H\dot{H}\dot{\phi}_{m}\nonumber\\
   && -\big(\kappa\dot{H}k^2+\big((\xi-2\sigma)H^2-(\xi+2\sigma)\dot{H}\big)\big(k^2-3a^2\dot{H}\big)\big)b_{t}\psi_{1} +\Big( \big(\kappa k^2+3(\xi+2\sigma)a^2\dot{H}\big)a^3\dot{H}\overline{\rho}_{m,n} \nonumber\\
   && -H^2(k^2-3a^2\dot{H})\big(4\sigma a^3\overline{\rho}_{m,n}+3(\xi+2\sigma)\overline{J}\overline{\rho}_{m,nn}\big) \Big)\frac{\kappa}{(\xi+2\sigma)a^4 k^2}\phi_{m} \Big).\label{phib-constraint-1529}
\end{eqnarray}
Substituting this constraint into the newly obtained action allows us to eliminate the variable $\phi_{h}$. Subsequently, we define two new variables to eliminate the coupling terms between the time derivatives of the remaining variables,
\begin{eqnarray}
  &&\psi_{2}\equiv \varphi_{b}+ \frac{(\xi+2\sigma) b_{t}^2}{2\big(1-(\xi+\sigma)b_{t}^2\big)H}\psi_{1} +\frac{3\kappa b_{t}a \dot{H} \overline{\rho}_{m,n}}{2\big(1-(\xi+\sigma) b_{t}^2\big)(k^2-3a^2\dot{H})H k^2}\phi_{m}, \label{psi2-912}
  \\
  &&\psi_{3}\equiv \psi_{1} -\frac{\kappa \overline{\rho}_{m,n}}{(\xi+2\sigma)b_{t}a k^2}\phi_{m}. \label{psi3-912}
\end{eqnarray}
Using these two new variables, we can eliminate $\varphi_{b}$ and $\psi_{1}$ from the action. The resulting action can be written as 
\begin{eqnarray}
  S^{(2)}_{s}=\int dtd^3x \left[ \frac{3(\xi+2\sigma)^2b_{t}^2 a^3}{4\kappa\big(1-(\xi+\sigma)b_{t}^2\big)}\dot{\psi}_{3}^2 +\frac{a^2\overline{\rho}_{m,n}k^2}{2\overline{J}(k^2-3 a^2\dot{H})}\left(\frac{\dot{\phi}_{m}}{k}\right)^2 +\frac{a F_{1}(t,k)}{\kappa b_{t}^2\dot{H}k^2+2 F_{1}(t,k)}(k\dot{\psi}_{2})^2 +\cdots \right],\label{action-effect-1161125}
\end{eqnarray}
where $F_{1}(t,k)=H^2\big(1-(\xi+\sigma)b_{t}^2\big)\big(k^2-3 a^2\dot{H}\big)$ and all non-kinetic terms are contained in ``$\cdots$''. The action~\eqref{action-effect-1161125} is an effective action containing only the dynamical variables $\psi_3$, $\phi_{m}$, and $\psi_{2}$, while the variables $\phi_{h}$, $\varphi_{h}$, $\phi_{b}$, $\varphi_{b}$, $\varphi_{m}$, and $\phi_{\ell}$ depend on them. From the structure of the kinetic terms in the action, one finds that the free of ghost instability requires
\begin{eqnarray}
  \frac{3(\xi+2\sigma)^2b_{t}^2 a^3}{4\kappa\big(1-(\xi+\sigma)b_{t}^2\big)}> 0,\label{ghost1-scalar-1161530}\\
  \frac{a^2\overline{\rho}_{m,n}k^2}{2\overline{J}(k^2-3 a^2\dot{H})}> 0,\label{ghost2-scalar-1161530}\\
  \frac{a F_{1}(t,k)}{\kappa b_{t}^2\dot{H}k^2+2 F_{1}(t,k)}> 0.\label{ghost3-scalar-1161530}
\end{eqnarray}
Using the assumes $\overline{\rho}_{m}>0$ and $\overline{p}_{m}>0$, the condition $\overline{\rho}_{m,n}>0$ can be obtained by Eq.~\eqref{pm-2127}. Since $a>0$, $\kappa>0$, $\overline{J}>0$, and $\dot{H}<0$, the condition~\eqref{ghost2-scalar-1161530} is satisfied, while the conditions~\eqref{ghost1-scalar-1161530} and~\eqref{ghost3-scalar-1161530} reduce to
\begin{eqnarray}
  && 1-(\xi+\sigma)b_{t}^2>0,\label{ghost4-scalar-1161630}\\
  && \frac{1}{k^2}>\frac{1}{a^2}\Big( \frac{\kappa b_{t}^2}{6\big(1-(\xi+\sigma)b_{t}^2\big)H^2}+\frac{1}{3\dot{H}} \Big).\label{ghost5-scalar-1161630}
\end{eqnarray}
It is not difficult to see that the condition~\eqref{ghost4-scalar-1161630} holds true, since $|\xi b_{t}^2|,|\sigma b_{t}^2|\ll 1$. According to the condition~\eqref{ghost5-scalar-1161630}, the absence of ghost instability in the scalar perturbations may depend on the wavenumber $k$, and thus be frequency-dependent. If $\tfrac{\kappa b_{t}^2}{2\big(1-(\xi+\sigma)b_{t}^2\big)H^2}+\tfrac{1}{\dot{H}}> 0$, the condition~\eqref{ghost5-scalar-1161630} fails for large $k$ (i.e., in the high-frequency/short-wavelength perturbation), leading to a ghost instability. A physically viable theory of gravity must avoid ghost instability at all perturbation frequencies. Therefore, to ensure ghost-free perturbations for all $k$, we must enforce the stronger requirement that $\tfrac{\kappa b_{t}^2}{2\big(1-(\xi+\sigma)b_{t}^2\big)H^2}+\tfrac{1}{\dot{H}}\le 0$. This inequality can be simplified to
\begin{eqnarray}
  \frac{1-(\xi+\sigma)b_{t}^2}{b_{t}^2}\ge -\frac{\kappa\dot{H}}{2 H^2}.\label{ghost-scalar-1282053}
\end{eqnarray} 
Once astronomical observations determine the values of $H^2$ and $\dot{H}$, substituting them into this equation will impose a constraint on the parameters $\xi$, $\sigma$, and $b_{t}$. Further applying the condition $|\xi b_{t}^2|,|\sigma b_{t}^2|\ll 1$ yields an upper bound for $b_{t}^2$: $b_{t}^2\lesssim-\tfrac{2 H^2}{\kappa \dot{H}}$.
\par
Following the same procedure, we analyze the stability of scalar perturbations in Fourier space under two distinct gauge conditions: (i) $\alpha=0$, $\varphi_{h}=0$, and (ii) $\alpha=0$, $\varphi_{b}=0$.
\par
\textbf{Gauge condition} \bm{$\alpha=0,\; \varphi_{h}=0$}
\par
After imposing the gauge condition $\alpha=0$, $\varphi_{h}=0$ and eliminating all variables without kinetic terms in the action~\eqref{action-second-1030}, we obtain an effective action in Fourier space
\begin{eqnarray}
  S^{(2)}_{s}=\int dtd^3x \left[ \frac{3(\xi+2\sigma)^2b_{t}^4 a^3}{4\kappa\big(1-(\xi+\sigma)b_{t}^2\big)}\dot{\psi}_{5}^2 +\frac{a^2\overline{\rho}_{m,n}k^2}{2\overline{J}(k^2-3 a^2\dot{H})}\left(\frac{\dot{\phi}_{m}}{k}\right)^2 +\frac{a F_{1}(t,k)}{\kappa b_{t}^2\dot{H}k^2+2 F_{1}(t,k)}(k\dot{\psi}_{7})^2 +\cdots \right].\label{effective-action-1171548}
\end{eqnarray}
Here, all non-kinetic terms are contained in ``$\cdots$''. During the derivation of this action, we successively introduce new variables
\begin{equation}
\begin{aligned}
  &\psi_{4} \equiv E +\frac{(\xi+2\sigma) b_{t}^2}{1-(\xi+\sigma) b_{t}^2}\phi_{h} -\frac{(\xi+2\sigma) b_{t}}{1-(\xi+\sigma) b_{t}^2}\phi_{b},\qquad \psi_{5} \equiv \phi_{h} -\frac{1}{b_{t}}\phi_{b},  \\
  & \psi_{6} \equiv \varphi_{b} - \frac{b_{t}}{2 H} \psi_{4},\qquad \psi_{7} \equiv \psi_{6} + \frac{\kappa b_{t}\overline{\rho}_{m,n}}{2a H\big(1-(\xi+\sigma) b_{t}^2\big) (k^2-3a^2\dot{H})}k^2 \phi_{m}.
\end{aligned}
\end{equation}
Comparing the actions in Eqs.~\eqref{action-effect-1161125} and action~\eqref{effective-action-1171548} shows that the scalar perturbations in the gauge $\alpha=0$, $E=0$ satisfy the same ghost-free conditions as those in the gauge $\alpha=0$, $\varphi_{h}=0$.
\par
\textbf{Gauge condition} \bm{$\alpha=0,\; \varphi_{b}=0$}
\par
Under the gauge condition $\alpha=0$, $\varphi_{b}=0$, we eliminate the non-dynamical variables from the action in Eq.~\eqref{action-second-1030} to obtain the following effective action in Fourier space
\begin{eqnarray}
  S^{(2)}_{s}=\int dtd^3x \left[ \frac{3(\xi+2\sigma)^2b_{t}^2 a^3}{4\kappa\big(1-(\xi+\sigma)b_{t}^2\big)}\dot{\psi}_{9}^2 +\frac{a^2\overline{\rho}_{m,n}k^2}{2\overline{J}(k^2-3 a^2\dot{H})}\left(\frac{\dot{\phi}_{m}}{k}\right)^2 +\frac{b_{t}^2}{4H^2}\frac{a F_{1}(t,k)}{\kappa b_{t}^2\dot{H}k^2+2 F_{1}(t,k)}(k\dot{\psi}_{8})^2 +\cdots \right].\label{effective-action-1282028}
\end{eqnarray}
Here, all non-kinetic terms are contained in ``$\cdots$''. In deriving this action, we successively introduce a series of new variables
\begin{eqnarray}
  \psi_{8} \equiv E +\frac{(\xi+2\sigma) b_{t}^2}{1-(\xi+\sigma) b_{t}^2}\phi_{h} -\frac{(\xi+2\sigma) b_{t}^2}{1-(\xi+\sigma) b_{t}^2}\phi_{b} -\frac{\kappa\overline{\rho}_{m,n}}{\big(1-(\xi+\sigma) b_{t}^2\big)\big(k^2-3a^2\dot{H}\big)a}\phi_{m}, \qquad \psi_{9} \equiv \phi_{b} -b_{t}\phi_{h}.
\end{eqnarray}
A comparison of this action~\eqref{effective-action-1282028} with the actions in Eqs.~\eqref{action-effect-1161125} and~\eqref{effective-action-1171548} reveals that its ghost-free conditions are identical in the gauge $\alpha=0$, $E=0$ and the gauge $\alpha=0$, $\varphi_{h}=0$.
\par
We have analyzed the stability of scalar perturbations under three different gauge choices and found identical results, as expected, given that gauge degrees of freedom do not affect physical observables. The effective actions in Eqs.~\eqref{action-effect-1161125}, \eqref{effective-action-1171548}, and \eqref{effective-action-1282028} contain only dynamical variables, which confirms that the Bumblebee model contains two scalar modes, in addition to the one in the matter sector. Regarding ghost instability, the condition (Eq.~\eqref{ghost5-scalar-1161630}) for its absence can, in principle, depend on frequency (or wavenumber). However, under the small-parameter assumption $|\xi b_{t}^2|,|\sigma b_{t}^2|\ll 1$, the condition~\eqref{ghost-scalar-1282053}, which reduces to $b_{t}^2\lesssim \tfrac{2 H^2}{-\kappa\dot{H}}$, ensures that the scalar perturbations are ghost-free at all frequencies.
\par
Note that an examination of the key Eqs.~\eqref{psi1-912}, \eqref{phib-constraint-1529}, \eqref{psi3-912}, \eqref{action-effect-1161125}, \eqref{effective-action-1171548}, and \eqref{effective-action-1282028} suggests that the aforementioned stability conditions may no longer hold under the parameter condition $\sigma=-\frac{1}{2}\xi$. However, under the condition $\sigma=-\frac{1}{2}\xi$, the background equation~\eqref{EqA-BG} reduces to $H^2=-2\kappa\overline{V}_{,b^2}/(3\xi)=constant$, which is inconsistent with astronomical observations. Consequently, this case is excluded from our analysis. It should be noted that a time-dependent background value $b_{t}$ does not generally yield a constant Hubble parameter under the condition $\sigma=-\frac{1}{2}\xi$. This point is discussed in Ref.~\cite{vandeBruck:2025aaa}, where it was shown that applying this condition results in only one scalar mode, in addition to the one present in the matter sector.

\subsection{The small-scale limit}
In the previous subsection, we studied the ghost-free conditions for the scalar perturbations in the Bumblebee model. We now turn to the stability and propagation characteristics of GWs in the small-scale limit ($k\rightarrow \infty$). Since the physics is gauge-invariant, we perform this analysis specifically in the gauge condition $\alpha=0, E=0$.
\par
In the small-scale limit, retaining only the $k^2$-order terms, the action~\eqref{action-effect-1161125} simplifies to
\begin{eqnarray}
  S^{(2)}_{s}\approx \int dtd^3x &\Bigg[& \frac{3(\xi+2\sigma)^2b_{t}^2 a^3}{4\kappa\big(1-(\xi+\sigma)b_{t}^2\big)}(\dot{\psi}_{3})^2 +\frac{a^2\overline{\rho}_{m,n}}{2\overline{J}}\left(\frac{\dot{\phi}_{m}}{k}\right)^2 -\frac{F_{2}(t)}{4\kappa}a k^2\psi_{3}^2 -\frac{\overline{\rho}_{m,nn}}{2 a^3}k^2\left(\frac{\phi_{m}}{k}\right)^2 \nonumber\\
  && +\frac{\big(1-(\xi+\sigma)b_{t}^2\big) a H^2}{2\big(1-(\xi+\sigma)b_{t}^2\big)H^2+\kappa b_{t}^2\dot{H}}(k\dot{\psi}_{2})^2 -\frac{\kappa(2-3\xi b_{t}^2)+2(\xi+2\sigma)\xi b_{t}^2}{\kappa \big(\kappa\dot{H}b_{t}^2+2\big(1-(\xi+\sigma)b_{t}^2\big)H^2\big)}a H^2 k\psi_{3}(k\dot{\psi}_{2}) \;\Bigg],\label{action-k2-1908}
\end{eqnarray}
where the specific form of $F_{2}(t)$ is
\begin{eqnarray}
  F_{2}(t)&=&\Big( -\big(1-(\xi+\sigma)b_{t}^2\big)\big(\kappa(2-3\xi b_{t}^2)^2+2b_{t}^2(\xi-6\sigma)(\xi+2\sigma)\big)\big(1-(2\xi+\sigma)b_{t}^2\big)H^2 \nonumber\\
  && +b_{t}^4 \dot{H}(\xi+2\sigma)^2\big(2\xi^2 b_{t}^2+\kappa\big(3-(4\xi+3\sigma)b_{t}^2\big)\big) \Big)/\Big( \big(1-(\xi+\sigma)b_{t}^2\big)^2\big(\kappa b_{t}^2\dot{H}+2\big(1-(\xi+\sigma)b_{t}^2\big)H^2\big) \Big).
\end{eqnarray}
In this action, it is $\dot{\psi}_{2}$, and not $\psi_{2}$ itself, that contributes to the $k^2$-order terms associated with the variable $\psi_{2}$. Varying the action~\eqref{action-k2-1908} with respect to $\psi_{2}$ yields the following equation of motion (analogous to the treatment of the vector perturbations)
\begin{eqnarray}
  k \dot{\psi}_{2}\approx \frac{\kappa(2-3\xi b_{t}^2)+2\xi b_{t}^2(\xi+2\sigma)}{2\kappa\big(1-(\xi+\sigma)b_{t}^2\big)} k \psi_{3}.\label{psidt-929}
\end{eqnarray}
Here, we consider only the dynamical effects and have accordingly simplified the expression. Substituting this constraint into the action~\eqref{action-k2-1908}, an effective action that contains only the dynamical variables is obtained
\begin{eqnarray}
  S^{(2)}_{s}\approx\int dtd^3x \left[ \frac{3(\xi+2\sigma )^2b_{t}^2a^3}{4\kappa\big(1-(\xi+\sigma)b_{t}^2\big)}\left( (\dot{\psi}_{3})^2 -c_s^2\frac{k^2}{a^2}\psi_{3}^2 \right) +\frac{a^2\overline{\rho}_{m,n}}{2\overline{J}}\left( \left(\frac{\dot{\phi}_{m}}{k}\right)^2 -\frac{\overline{p}_{m,n}}{\overline{\rho}_{m,n}}\frac{k^2}{a^2}\left(\frac{\phi_{m}}{k}\right)^2 \right) \right],\label{smallLimit-S2-1001}
\end{eqnarray}
where $c_s^2\equiv\big(3\kappa+2\xi b_{t}^2(\xi-2\kappa)-3\kappa\sigma b_{t}^2\big)/\big(3\kappa(1-(\xi+\sigma) b_{t}^2)\big)$. It is straightforward to see that the variables $\phi_{h}$, $\varphi_{h}$, $\phi_{b}$, $\varphi_{b}$, $\varphi_{m}$, and $\phi_{\ell}$ depend on $\psi_{2}$, $\psi_{3}$, and $\phi_{m}$, as shown in Eqs.~(\ref{phiell-912})-(\ref{psi3-912}). The action~\eqref{action-k2-1908} indicates that $\psi_{2}$ is a cyclic coordinate in the small-scale limit, and the constraint~\eqref{psidt-929} implies that $\dot{\psi}_{2}$ depends on $\psi_{3}$.
\par
The action~\eqref{smallLimit-S2-1001} is an effective action that contains only the dynamical variables $\psi_{3}$ and $\phi_{m}$. The free of ghost instability in the small-scale limit requires
\begin{eqnarray}
  \frac{3(\xi+2\sigma )^2b_{t}^2a^3}{4\kappa\big(1-(\xi+\sigma)b_{t}^2\big)}&>&0, \\
  \frac{a^2}{2\overline{J}}\overline{\rho}_{m,n}&>&0.
\end{eqnarray}
Since $|\xi b_{t}^2|,|\sigma b_{t}^2|\ll 1$, $a>0$, and $\kappa>0$, the perturbation $\psi_{3}$ is free of ghost instability in the small-scale limit. For the variable $\phi_{m}$, since $\overline{J}>0$ and $\overline{\rho}_{m,n}>0$, it is also ghost-free in the small-scale limit.
\par
For the Laplacian stability, it requires
\begin{eqnarray}
  \frac{3\kappa+2\xi b_{t}^2(\xi-2\kappa)-3\kappa\sigma b_{t}^2}{3\kappa(1-(\xi+\sigma) b_{t}^2)}&>&0,\label{cs2-2137}\\
  \frac{\overline{p}_{m,n}}{\overline{\rho}_{m,n}}&>&0.\label{cm2-2137}
\end{eqnarray}
The first condition~\eqref{cs2-2137} is obviously satisfied since $|\xi b_{t}^2|,|\sigma b_{t}^2|\ll 1$, as will be shown in Eq.~\eqref{cs2-expand-2138} below. Furthermore, the condition~\eqref{cm2-2137} implies $\overline{p}_{m,n}>0$. This means that the greater the particle number density, the higher the pressure $\overline{p}_{m}$ contributed by the matter, which is a result consistent with physical intuition.
\par
Varying the action~\eqref{smallLimit-S2-1001} with respect to $\psi_{3}$ and $\phi_{m}$, respectively, their dispersion relations in the small-scale limit are
\begin{eqnarray}
   w_{\psi_3}^2 -c_{s}^{2}\overline{g}^{ij}k_{i}k_{j}&=&0,\label{scalar-motion-9252231}\\
   w_{\phi_m}^2 -\frac{\overline{p}_{m,n}}{\overline{\rho}_{m,n}}\overline{g}^{ij}k_{i}k_{j}&=&0.
\end{eqnarray}
Here, $\overline{p}_{m,n}/\overline{\rho}_{m,n}$ is the matter sound speed squared. The propagation speed squared $c_s^2$ of the scalar perturbations is
\begin{eqnarray}
  c_s^2=\frac{3\kappa+2\xi b_{t}^2(\xi-2\kappa)-3\kappa\sigma b_{t}^2}{3\kappa(1-(\xi+\sigma) b_{t}^2)}=1+\frac{\kappa-2\xi}{3\kappa}(-\xi b_{t}^2)+\mathcal{O}\big((\xi b_t^2)^n(\sigma b_t^2)^{m}\big),\label{cs2-expand-2138}
\end{eqnarray}
where $n+m\ge 2$. Avoiding tachyonic instability in the vector perturbations requires $\xi\le 0$, as shown in Eq.~\eqref{Tachyon-vector-1736}. Consequently, the propagation speed of the scalar perturbations exceeds 1, which results from the violation of Lorentz symmetry. If the Lorentz-violating parameter $\xi b_{t}^2$ equals zero, the speed will return to 1.
\par
In conclusion, in the small-scale limit, the matter sound speed is given by $\sqrt{\overline{p}_{m,n}/\overline{\rho}_{m,n}}$. The propagation speed of scalar GWs is greater than 1, a direct consequence of Lorentz violation through the parameter $\xi b_{t}^2$. Indeed, this superluminality vanishes when $\xi b_{t}^2=0$, in which case the scalar GW speed equals 1 for any $\sigma$. Finally, the scalar modes in the Bumblebee model are free of both ghost and Laplacian instabilities, provided that $b_{t}^2\lesssim \tfrac{2 H^2}{-\kappa\dot{H}}$, $|\xi b_{t}^2|,|\sigma b_{t}^2|\ll 1$, and $\overline{p}_{m,n}>0$.

\section{Polarization modes of gravitational waves} \label{polarization-section-914013}
\begin{figure}[H]
	\centering
  \subfloat[$P_{l}$: longitudinal mode]{
        \includegraphics[width=0.3\textwidth]{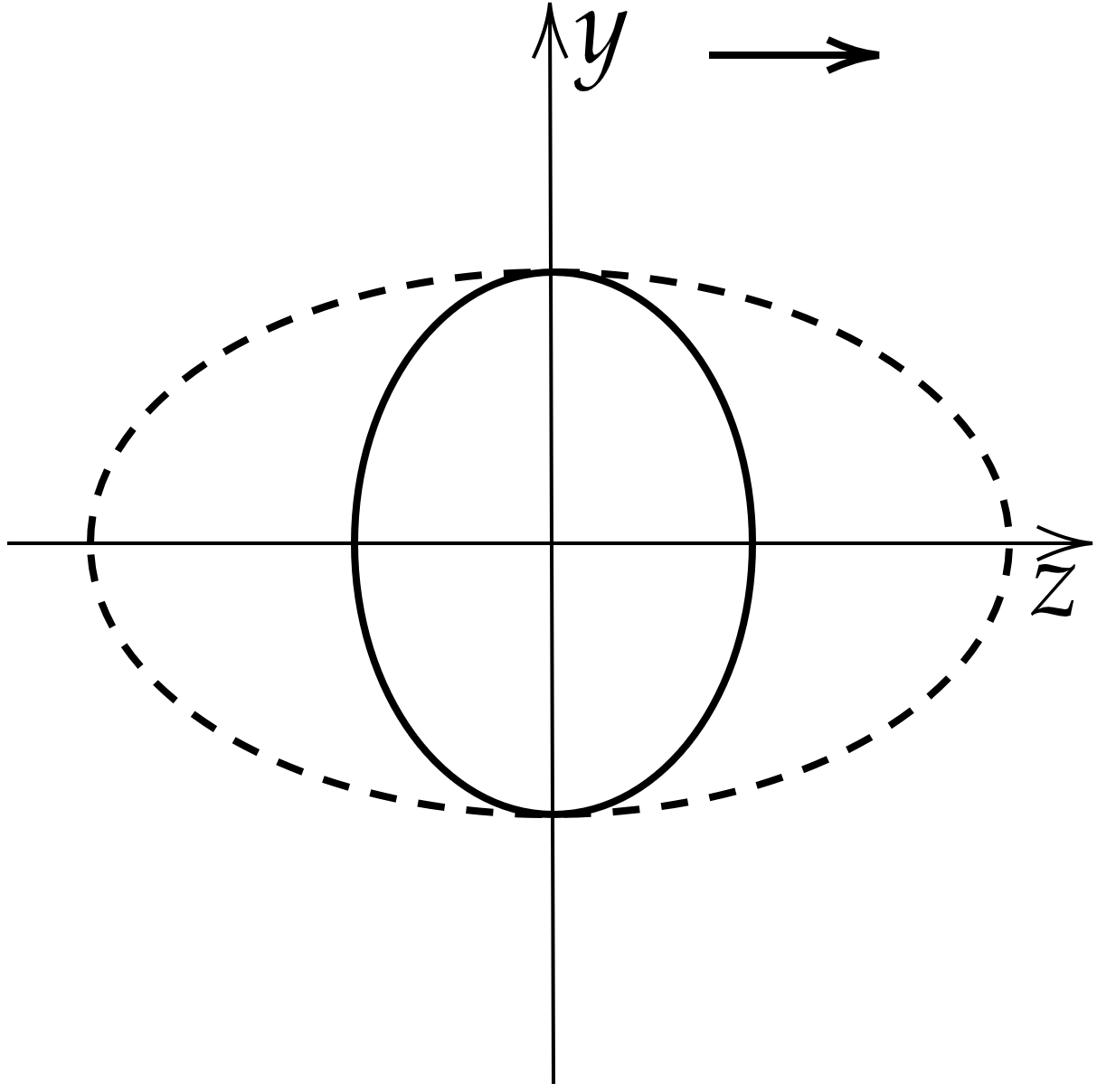}
    }
	\subfloat[$P_{x}$: vector-$x$ mode]{
        \includegraphics[width=0.3\textwidth]{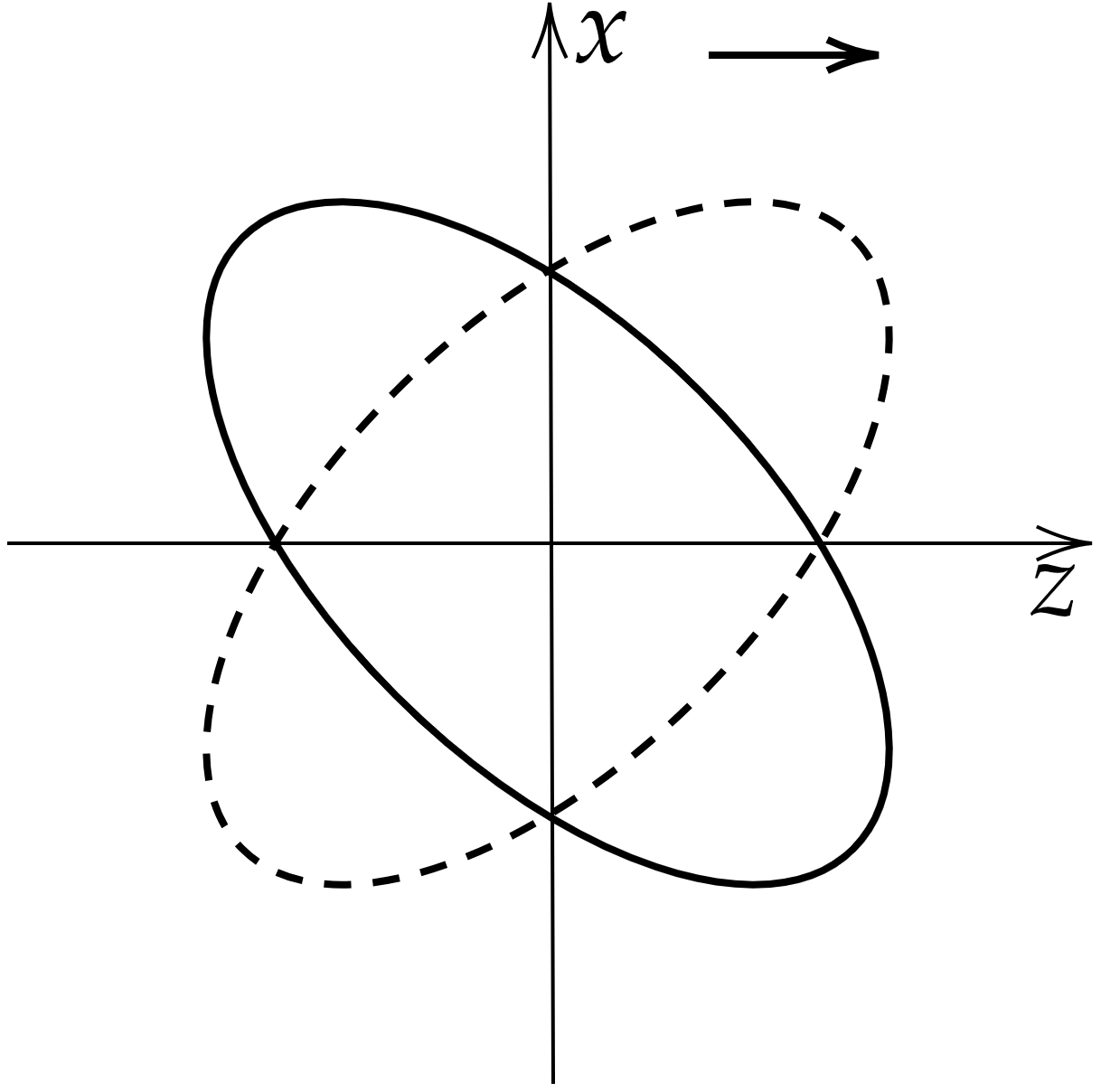}
    }
	\subfloat[$P_{y}$: vector-$y$ mode]{
        \includegraphics[width=0.3\textwidth]{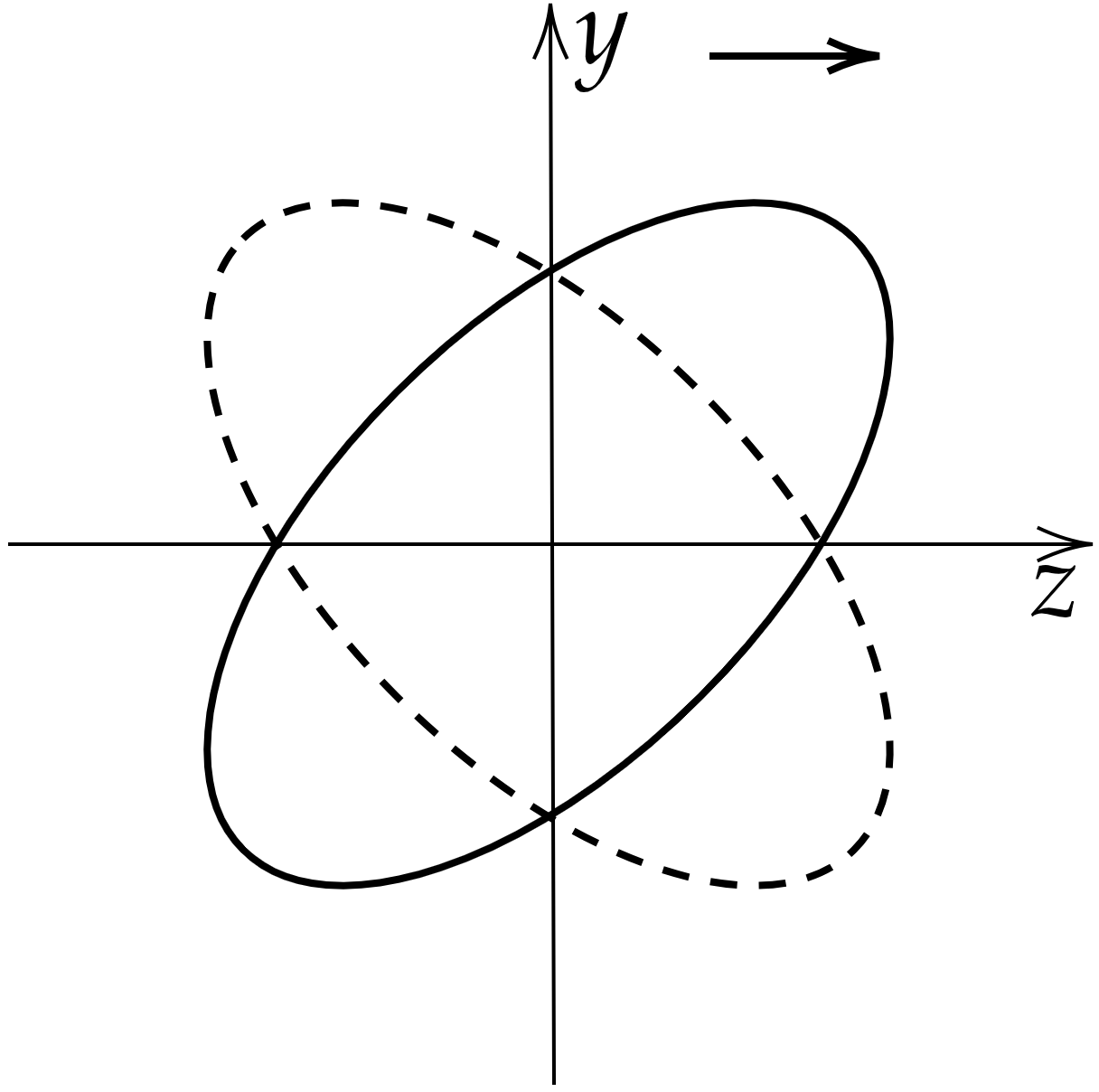}
    }\\
	\subfloat[$P_{+}$: plus mode]{
        \includegraphics[width=0.3\textwidth]{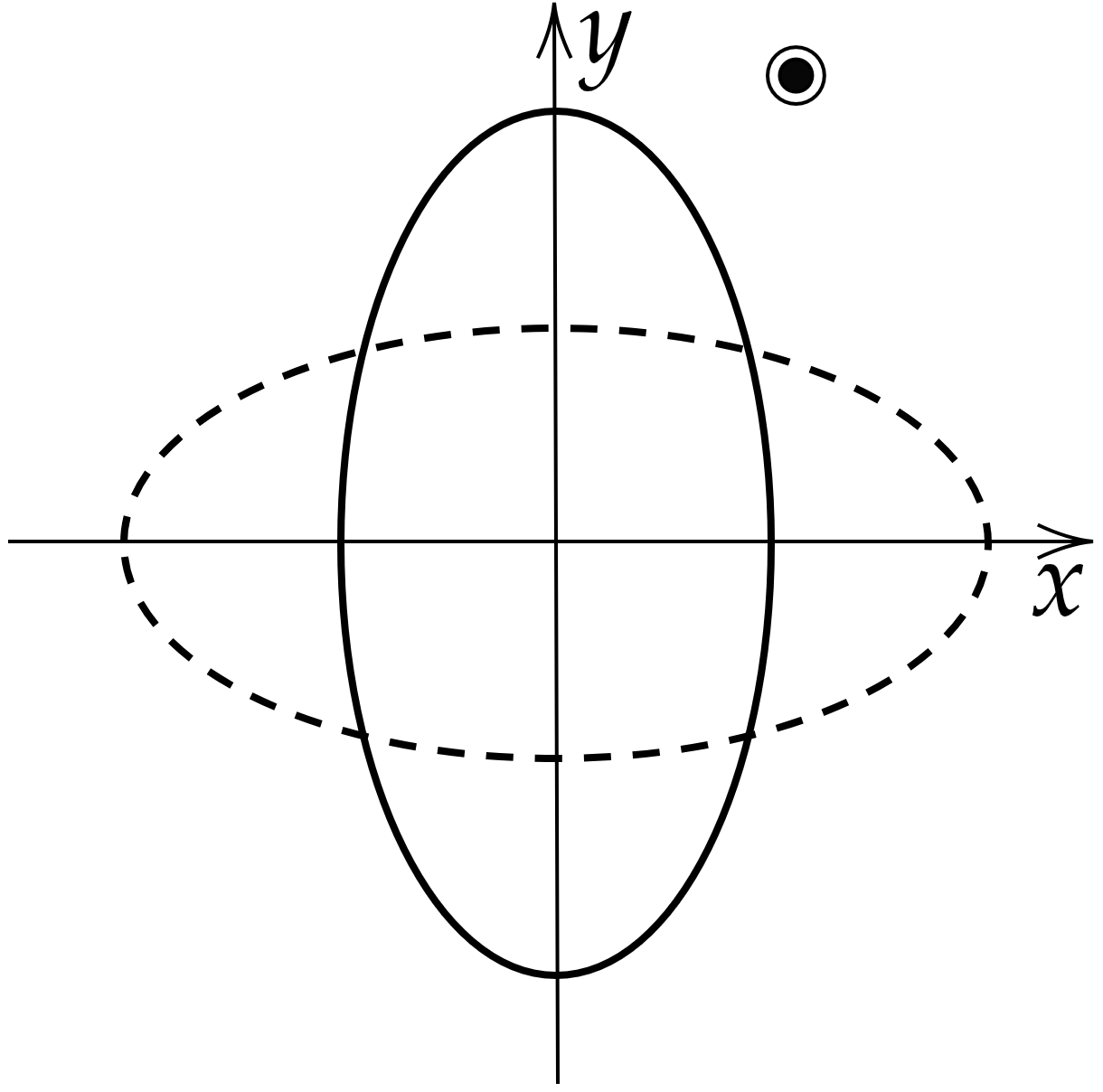}
    }
	\subfloat[$P_{\times}$: cross mode]{
        \includegraphics[width=0.3\textwidth]{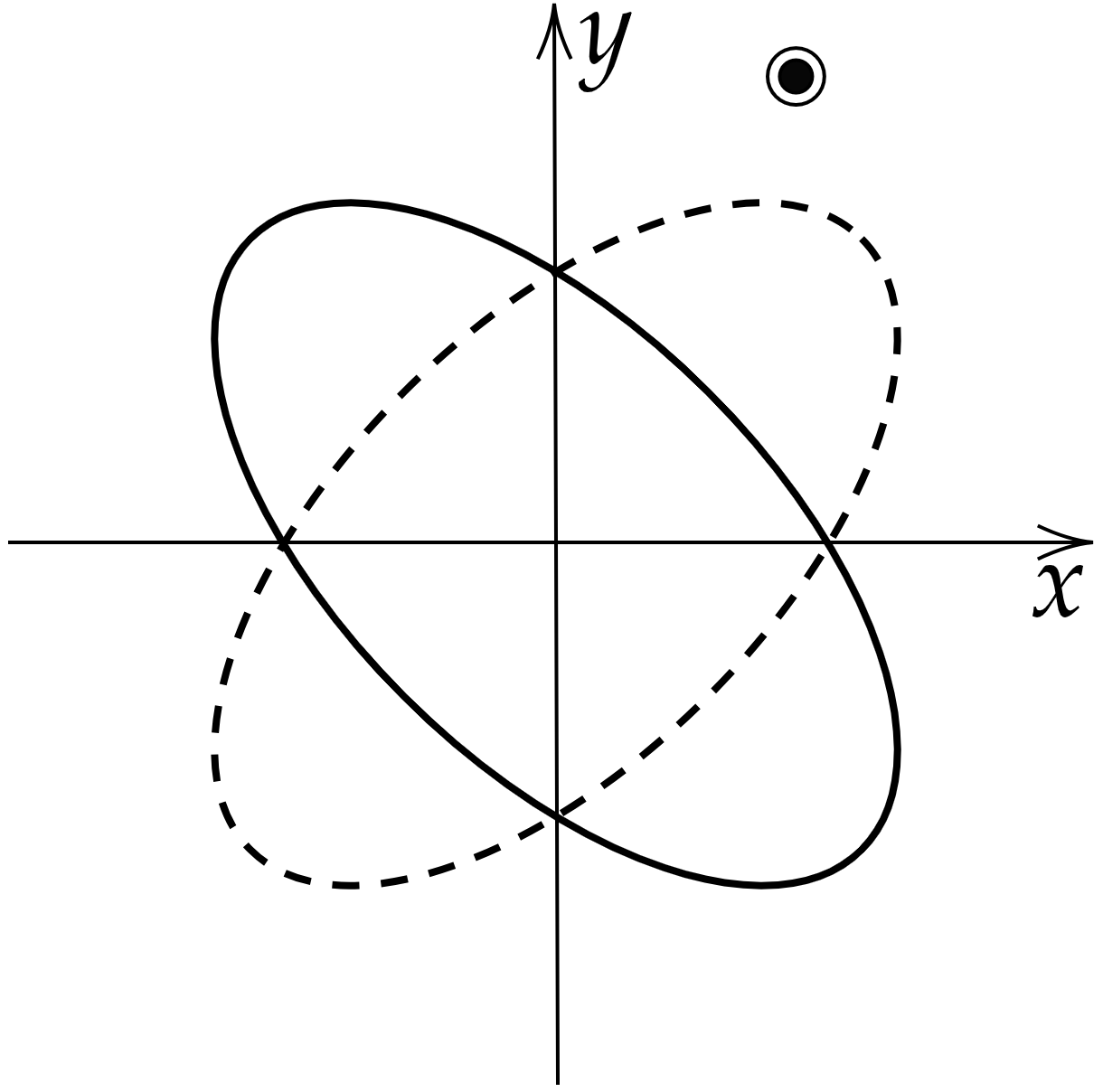}
    }
	\subfloat[$P_{b}$: breathing mode]{
        \includegraphics[width=0.3\textwidth]{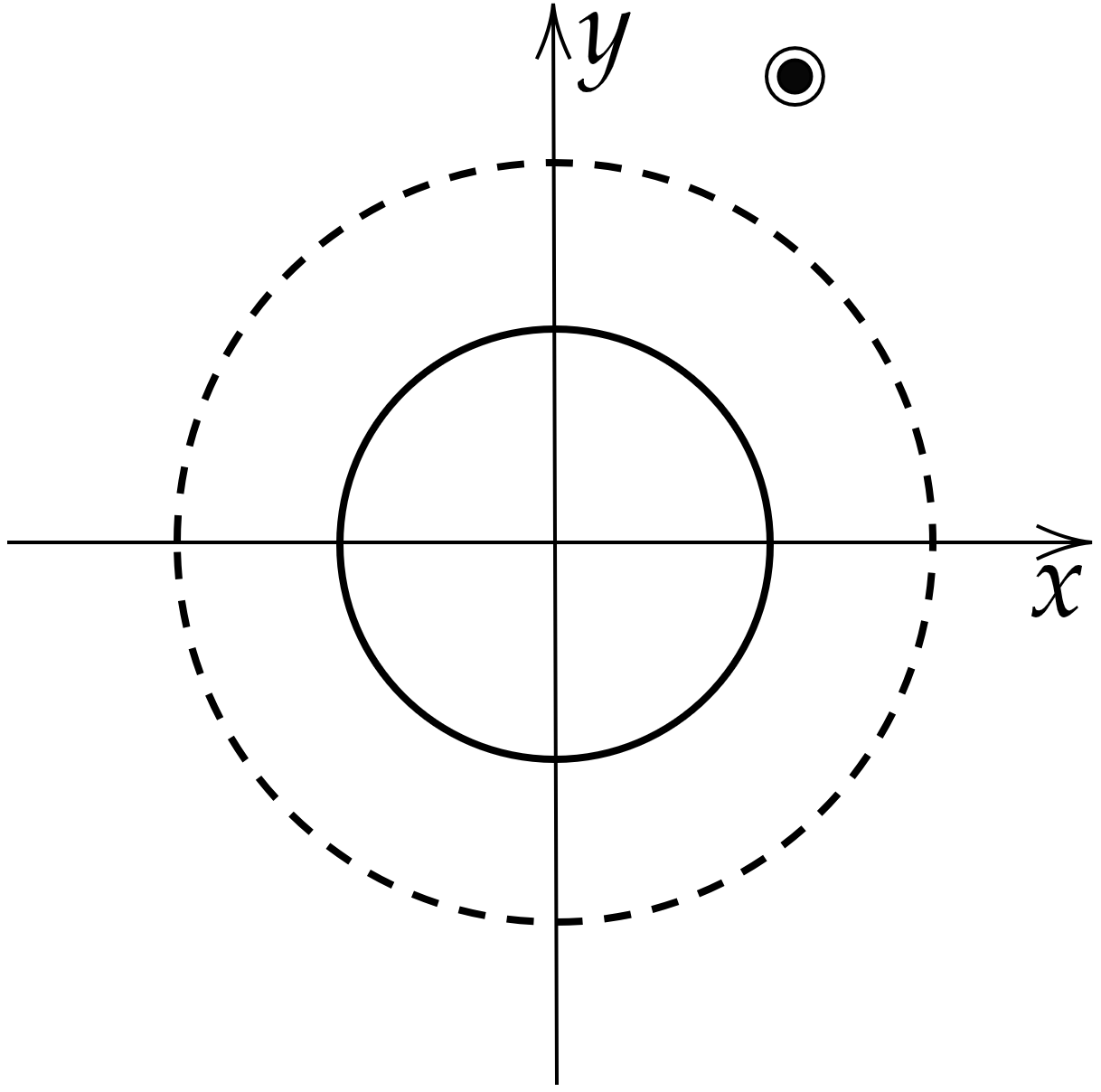}
    }\\	
    \caption{This figure illustrates the six polarization modes of GWs~\cite{Eardley:1973zuo}. Assuming the propagation direction of GWs is along the ``$+z$" direction, these panels show the variation in the relative positions of test particles distributed on a spherical surface. No relative displacement occurs for test particles located along the direction perpendicular to the plane of each figure. The solid and dashed lines represent a phase difference of $\pi$, where the symbol ``$\rightarrow$" denotes wave propagation to the right, and ``$\odot$" indicates the wave propagation outward from the page.}
    \label{polarization-diagrams-9122325}
\end{figure}

In this section, we study the polarization modes of GWs in the Bumblebee model under the limit $a\rightarrow 1, \dot{a}\rightarrow 0, \ddot{a} \rightarrow 0$.
According to the general principle of relativity, GWs cannot be identified by detecting the motion of a single particle. Instead, they can be detected by observing the relative displacement between two adjacent free test particles. This relative displacement, which is described by the geodesic deviation equation, provides a specific manifestation of the polarizations of GWs
\begin{eqnarray}
  \frac{\mathrm{D}^2(\delta x^{\mu})}{\mathrm{d}\tau^2}=-R^{\mu}_{\;\;\rho\nu\lambda}\frac{\mathrm{d}x^{\rho}}{\mathrm{d}\tau}\frac{\mathrm{d}x^{\lambda}}{\mathrm{d}\tau}\delta x^{\nu}.
\end{eqnarray}
We will focus on the background metric~\eqref{FRW-metric}, which implies a comoving coordinate system. Before the arrival of the GWs, the geodesic deviation equation can be simplified
\begin{eqnarray}
  \frac{\mathrm{D}^2(\delta x^{\mu})}{\mathrm{d}t^2}=-\overline{R}^{\mu}_{\;\; t\nu t}\delta x^{\nu}=\frac{\ddot{a}}{a}(\delta^{\mu}_{\;\; i}\delta^{i}_{\;\;\nu})\delta x^{\nu}.
\end{eqnarray}
It is easy to see that a relative acceleration exists between any two free test particles, even in the absence of GWs. This effect is caused by the accelerated expansion of the universe. For the GW detectors, this relative acceleration is very small. In this section, since we focus on the polarization of GWs, we will neglect this effect. Assuming that GWs are sufficiently weak, it is reasonable to adopt the following approximations: $\tau\approx t$, $\delta x^{t}\approx 0$, and $\mathrm{d}x^{\mu}/\mathrm{d}\tau\approx (1,0,0,0)$. When GWs pass by, the relative acceleration between any two free test particles is given by
\begin{eqnarray}
  a_{\mu}\equiv g_{\mu\nu}\frac{\mathrm{D}^2(\delta x^{\nu})}{\mathrm{d}\tau^2} \approx -R^{(1)}_{\;\;\; \mu 0 j 0}\delta x^{j}. \label{aAcceleration-2247}
\end{eqnarray}
By calculations, it is not difficult to find that $R^{(1)}_{\;\;\; 0 0 i 0}=0$. Thus, the relative accelerations between free test particles are primarily determined by the components of $R^{(1)}_{\;\;\; i 0 j 0}$. Considering the gauge conditions $\varepsilon_{i}=0, \alpha=0, \varphi_{h}=0$ and the coordinate system where GWs propagate along the ``$+z$'' direction, the specific form of $R^{(1)}_{\;\;\; i 0 j 0}$ is given by
\begin{eqnarray}
  R^{(1)}_{\;\;\; i 0 j 0}&=&-\frac{a}{2}
  \left(
  \begin{array}{ccc}
    \makecell[c]{\Big( \partial_{t}^2(a E)+\ddot{a}E-2\dot{a}\dot{\phi}_{h} \\ +\big(\partial_{t}^2(a h_{+})+\ddot{a}h_{+}\big) \Big)} & \partial_{t}^2(a h_{\times})+\ddot{a}h_{\times} & -\frac{1}{a}\partial_{z}\dot{\lambda}_{x} 
    \\
    \partial_{t}^2(a h_{\times})+\ddot{a}h_{\times} & \makecell[c]{\Big( \partial_{t}^2(a E)+\ddot{a}E-2\dot{a}\dot{\phi}_{h} \\ -\big(\partial_{t}^2(a h_{+})+\ddot{a}h_{+}\big) \Big)} & -\frac{1}{a}\partial_{z}\dot{\lambda}_{y}
    \\
    -\frac{1}{a}\partial_{z}\dot{\lambda}_{x} & -\frac{1}{a}\partial_{z}\dot{\lambda}_{y} & \Big(\partial_{t}^2(aE)+\ddot{a}E-2\dot{a}\dot{\phi}_{h}-\frac{2}{a}\partial^2_{z}\phi_{h}\Big)
  \end{array}
  \right) \nonumber
  \\
  &=&-\frac{a}{2}
  \left(
  \begin{array}{ccc}
    (P_{b}+P_{+}) & P_{\times} &  P_{x}
    \\
    P_{\times} & (P_{b}-P_{+}) & P_{y}
    \\
    P_{x} & P_{y} & P_{l}
  \end{array}
  \right) \nonumber
  \\
  &\approx& -\frac{1}{2}
  \left(
  \begin{array}{ccc}
    \big(\ddot{E}+\ddot{h}_{+}\big) & \ddot{h}_{\times} & -\partial_{z}\dot{\lambda}_{x} 
    \\
    \ddot{h}_{\times} & \big(\ddot{E}-\ddot{h}_{+}\big) & -\partial_{z}\dot{\lambda}_{y}
    \\
    -\partial_{z}\dot{\lambda}_{x} & -\partial_{z}\dot{\lambda}_{y} & \big(\ddot{E}-2\partial^2_{z}\phi_{h}\big)
  \end{array}
  \right).\label{Ra-polarization-2302}
\end{eqnarray}
Here, $P_{+}$, $P_{\times}$, $P_{x}$, $P_{y}$, $P_{b}$, and $P_{l}$ represent the plus, cross, vector-$x$, vector-$y$, breathing, and longitudinal modes, respectively. Equation~\eqref{Ra-polarization-2302} reveals that the expansion (or contraction) of the universe also contributes to the polarization effects of GWs. However, this contribution is negligible compared to that of the GWs themselves for detector applications, and is therefore omitted. The right-hand side of the ``$\approx$'' symbol is obtained by taking the limit $a\rightarrow 1, \dot{a}\rightarrow 0, \ddot{a} \rightarrow 0$. Substituting these components into Eq.~\eqref{aAcceleration-2247} yields the schematic diagrams of the polarization modes shown in Fig.~\ref{polarization-diagrams-9122325}. These modes can be straightforwardly derived by considering a monochromatic plane wave.
\par
Taking the limit $a\rightarrow 1, \dot{a}\rightarrow 0, \ddot{a} \rightarrow 0$ into the background equations~(\ref{EqN-BG})-(\ref{EqA-BG}), we can obtain these constraints
\begin{eqnarray}
  \overline{V}_{,b^2}=0, \qquad \overline{\rho}_{m,n}=0, \qquad \overline{\rho}_{m}+\overline{V}+\frac{\Lambda}{\kappa}=0, \label{constraint-BGa1-2143}
\end{eqnarray}
where $b_{t}^2 \ne 0$ has been used. Obviously, the background potential $\overline{V}$ is determined by the background energy density $\overline{\rho}_{m}$ and the cosmological constant $\Lambda$. Furthermore, the constraint $\overline{V}_{,b^2}=0$ indicates that the potential $V(B^{\mu}B_{\mu}+b^2)$ contains no linear term in $B^{\mu}B_{\mu}+b^2$.
\par
\textbf{For the tensor perturbations:} Submitting the limit $a\rightarrow 1, \dot{a}\rightarrow 0, \ddot{a} \rightarrow 0$ into Eq.~\eqref{motion-hb-2340}, the equation of motion can be simplified
\begin{eqnarray}
  \ddot{h}_w-\frac{1-\sigma b_{t}^2}{1-\sigma b_t^2-\xi b_t^2}\partial_z\partial_z h_w=0.\label{tensor-mode-1191254}
\end{eqnarray}
According to the relation~\eqref{Ra-polarization-2302} between the polarization modes and the components of the Riemann tensor, $P_{+}\approx \ddot{h}_{+}$ and $P_{\times}\approx \ddot{h}_{\times}$, the two independent tensor modes ($P_{+}$ and $P_{\times}$) are permitted to propagate. Since tachyonic stability requires $\xi\le 0$~\eqref{Tachyon-vector-1736}, the speed of the tensor modes is less than the speed of light, with $c_{t}=\sqrt{(1-\sigma b_{t}^2)/(1-\sigma b_t^2-\xi b_t^2)}\le 1$. The equality holds only when the Lorentz-violating parameter $\xi b_{t}^2$ is zero.
\par
\textbf{For the vector perturbations:} We will consider the gauge condition $\varepsilon_{i}=0$. In Fourier space, taking the limit $a\rightarrow 1, \dot{a}\rightarrow 0, \ddot{a} \rightarrow 0$, the equations of motion for vector modes can be derived from Eqs.~(\ref{motion1-vector-2117})-(\ref{motion4-vector-2117})
\begin{eqnarray}
  &&\lambda_{p}=\frac{\xi b_{t}}{1-(\xi+\sigma) b_{t}^2}\zeta_{p},\label{lambda-polarization-1306}\\
  &&\ddot{\zeta}_{p}+\left( 1+\frac{\xi^2 b_{t}^2}{2\kappa\big(1-(\xi+\sigma) b_{t}^2\big)} \right) \delta^{ij}k_{i}k_{j}\zeta_{p}=0.\label{vector-mode-9261534}
\end{eqnarray}
According to the relation~\eqref{Ra-polarization-2302}, $P_{x}\approx -\partial_{z}\dot{\lambda}_{x}$ and $P_{y}\approx -\partial_{z}\dot{\lambda}_{y}$, there are two independent vector modes ($P_{x}$ and $P_{y}$) in the Bumblebee model. Since the Lorentz-violating parameters are very small $|\xi b_{t}^2|,|\sigma b_{t}^2|\ll 1$, the speed of the vector modes exceeds the speed of light. However, if $\xi b_{t}^2=0$ (where the propagation speed of the vector perturbations equals the speed of light), GWs of the vector modes are prohibited from propagating, as shown in Eq.~\eqref{lambda-polarization-1306}.
\par
Returning to Sec.~\ref{gaugeSec-vector-2109}, these results for the vector modes differ significantly from the predictions of Eq.~\eqref{motion-vector-9261532}. However, taking the limit $a\rightarrow 1, \dot{a}\rightarrow 0, \ddot{a} \rightarrow 0$ in Eq.~\eqref{motion-vector-9261532} yields the same result as Eq.~\eqref{vector-mode-9261534}.
\par
\textbf{For the scalar perturbations:} We will consider the gauge condition $\alpha=0$ and $\varphi_{h}=0$. In Fourier space, substituting the limit $a\rightarrow 1, \dot{a}\rightarrow 0, \ddot{a} \rightarrow 0$, and the constraint~\eqref{constraint-BGa1-2143} into the scalar perturbation equations~(\ref{equation-scalar-phil-2147}), we obtain the equations of motion for scalar modes
\begin{eqnarray}
  && \phi_{h}=\frac{\kappa+2\xi b_{t}^2(\xi-\kappa)-\kappa \sigma b_{t}^2}{2\kappa\big(1-(\xi+\sigma) b_{t}^2\big)}E,  \label{scalar-phi-2226}\\
  && \ddot{E}+\frac{3\kappa+2\xi b_{t}^2(\xi-2\kappa)-3\kappa\sigma b_{t}^2}{3\kappa\big(1-(\xi+\sigma) b_{t}^2\big)}\delta^{ij}k_{i}k_{j}E+\frac{8\kappa}{3(\xi+2\sigma)^2}\big(1-(\xi+\sigma) b_{t}^2\big)\overline{V}_{,b^2b^2}E=0.\label{scalar-equation-lambda-2243}
\end{eqnarray}
The relations between the polarization modes and the scalar perturbations are given by $P_{b}\approx \ddot{E}$ and $ P_{l}\approx \ddot{E}-2\partial^2_{z}\phi_{h}$. Obviously, only one scalar mode is permitted: a mixed mode of $P_{b}$ and $P_{l}$. According to Eq.~\eqref{cs2-expand-2138}, the propagation speed of the scalar mode exceeds the speed of light. Equation~\eqref{scalar-equation-lambda-2243} shows that the scalar mode of GWs is massive, $m_{s}=\sqrt{8\kappa\big(1-(\xi+\sigma) b_{t}^2\big)\overline{V}_{,b^2b^2}/\big(3(\xi+2\sigma)^2\big)}$. 
\par
Revisiting Sec.~\ref{Gauge1-scalar-9252232}, these results about the scalar mode differ significantly from the predictions of Eq.~\eqref{scalar-motion-9252231}. This is because Eq.~\eqref{scalar-motion-9252231} represents the case of the small-scale limit. If the same limit is taken, the mass term in Eq.~\eqref{scalar-equation-lambda-2243} can be neglected, in which case the amplitude of the $P_{l}$ mode is much smaller than that of the $P_{b}$ mode,
\begin{eqnarray}
  P_{l}\approx \ddot{E}-2\partial^2_{z}\phi_{h}=\ddot{E}+2\frac{\kappa+2\xi^2b_{t}^2-\kappa b_{t}^2(2\xi+\sigma)}{2\kappa\big(1-(\xi+\sigma) b_{t}^2\big)}\delta^{ij}k_{i}k_{j}E=\mathcal{O}\big((\xi b_{t}^2)(\sigma b_{t}^2)^0\big)\delta^{ij}k_{i}k_{j}E=\mathcal{O}\big((\xi b_{t}^2)(\sigma b_{t}^2)^0\big)P_{b}. 
\end{eqnarray}
Here, Eq.~\eqref{scalar-equation-lambda-2243} has been used in the right-hand side of the third equality. According to Eq.~\eqref{scalar-equation-lambda-2243}, it follows that the same result is obtained when $\overline{V}_{,b^2b^2}=0$ even if the small-scale limit is not taken.
\par
From the scalar perturbation equations~(\ref{equation-scalar-phil-2147}) and the equations of motion~\eqref{scalar-phi-2226} and~\eqref{scalar-equation-lambda-2243}, the properties of the scalar GWs depend on the parameter space as follows:
\begin{itemize}
  \item Case 1: $\xi b_{t}^2\ne 0$. A single massive scalar mode exists, corresponding to the mixed mode of the $P_{b}$ and $P_{l}$. Especially, when $\overline{V}_{,b^2b^2}=0$, this scalar mode is massless and superluminal.
  \item Case 2: $\xi=0,\sigma b_{t}^2\ne 0$. A single massive scalar mode is allowed, corresponding to the mixed mode of the $P_{b}$ and $P_{l}$. Especially, when $\overline{V}_{,b^2b^2}=0$, this reduces to a massless $P_{b}$ mode propagating luminally.
  \item Case 3: $\xi b_{t}^2=\sigma b_{t}^2=0$. No scalar GWs exist in this regime. In particular, when $\xi=\sigma=0,b_{t}\ne 0,\overline{V}_{,b^2b^2}\ne 0$, there is no scalar perturbation degree of freedom in the Bumblebee model. In all other situations satisfying the case condition, a single scalar degree of freedom remains, originating from the Bumblebee field.
\end{itemize}
\par
Prior to this work, a study on the polarization modes of the Bumblebee theory was conducted in Ref.~\cite{Liang:2022hxd} where $\sigma=0$. The authors considered GWs propagating along the ``$+z$” direction with an arbitrary background vector field $b_{\mu}$ in Minkowski background. Their results indicate that the polarization modes of GWs depend on the parameter space, which is divided into three cases: (I) $b_{x}^2+b_{y}^2\ne 0$ and $b^{t}\ne b^{z}$; (II) $b_{x}^2+b_{y}^2= 0$ and $b^{t}\ne b^{z}$; (III) $b^{t}=b^{z}$. In Case II, which is consistent with our background conditions, five independent modes are presented: $P_{+}$, $P_{\times}$, $P_{x}$, and $P_{y}$, along with a massive mixture mode of the $P_{b}$ and $P_{l}$. Our results are consistent with these findings. The distinction lies in the fact that we determine a specific range for the GW speed, which agrees with the results from cosmological perturbations discussed above.
\par
In conclusion, the scale factor $a(t)$ influences the polarizations of GWs, but this effect is very weak compared to that of the perturbations in the cosmological background. In the limit $a\rightarrow 1, \dot{a}\rightarrow 0, \ddot{a} \rightarrow 0$, we find that the Bumblebee model allows the propagation of two tensor modes ($P_{+},P_{\times}$), two vector modes ($P_{x},P_{y}$), and one scalar mode (the $P_{b}$ or the mixture of the $P_{b}$ and $P_{l}$). All GWs of these modes propagate at speeds different from the speed of light when $\xi b_{t}^2\ne 0$. The speed of the tensor modes is less than the speed of light, while the speeds of the vector and scalar modes exceed it. For the mixed scalar mode of $P_{b}$ and $P_{l}$, GWs are massive. However, under either the small-scale limit or $\overline{V}_{,b^2b^2}=0$, the mass term for GWs vanishes. In these regimes, the amplitude of $P_{l}$ becomes significantly smaller than that of $P_{b}$. Furthermore, $P_{l}$ vanishes entirely if $\xi=0,\sigma b_{t}^2\ne 0$. In particular, when the Lorentz-violating parameters are zero ($\xi b_{t}^2=\sigma b_{t}^2=0$), all vector and scalar modes are non-propagating. In this case, only the tensor modes ($P_{+}$ and $P_{\times}$) propagate at the speed of light. We summarize these results in Table~\ref{Table-polarization-1108}. The condition $\xi\le 0$ is obtained by the tachyonic stability of the vector perturbations.
%
\begin{table}[h]
\centering
\begin{tabular}{|c|c|c|c|c|c|c|}
  \hline
  \textbf{Case} & \textbf{Conditions} & \textbf{Type} & \textbf{Modes} & \textbf{d.o.f.} & \textbf{Speed} & \textbf{Mass}
  \\ \hline
    & & Tensor & $P_{+},P_{\times}$ & 2 & $< 1$ & $0$
  \\ \cline{3-7}
  Case 1 & $\xi b_{t}^2\ne 0$ & Vector & $P_{x}, P_{y}$ & 2 & $> 1$ & $0$ 
  \\ \cline{3-7}
   & & Scalar & Mixture & 1 & $> 1$ & $\ne 0$
  \\ \hline
  & & Tensor & $P_{+},P_{\times}$ & 2 & $1$ & $0$  
  \\ \cline{3-7}
  Case 2 & $\xi=0,\sigma b_{t}^2\ne 0$ & Vector & no & 0 & - & - 
  \\ \cline{3-7}
  & & Scalar & Mixture/$P_{b}$ & 1 & $1$ & $\ne 0$
   \\ \hline
  &  & Tensor & $P_{+},P_{\times}$ & 2 & $1$ & $0$ 
  \\ \cline{3-7}
  Case 3 & $\xi b_{t}^2=\sigma b_{t}^2=0$ & Vector & no & 0 & - & -
  \\ \cline{3-7}
  & & Scalar & no & 0 & - & -
  \\ \hline
\end{tabular}
\caption{The polarization modes of GWs in the Bumblebee model. Here, we take the limit $a\rightarrow 1, \dot{a}\rightarrow 0, \ddot{a} \rightarrow 0$ and use the condition $\xi\le 0$. ``d.o.f.'' represents the degrees of freedom in the propagation of GWs. The speeds of the scalar modes represent the speed of the scalar GWs when the mass term is ignored. In case 2, the $P_{l}$ component within the mixed mode of the $P_{b}$ and $P_{l}$ vanishes if $\overline{V}_{,b^2b^2}=0$.}
\label{Table-polarization-1108}
\end{table}

\section{Conclusion} \label{conclusion-section-914018}
The formulation of Einstein's GR marks a qualitative leap in humanity's understanding of gravitational phenomena. Over the subsequent century, the exploration of gravity has continued unremittingly. In 2015, humanity achieved the first direct detection of GWs~\cite{LIGOScientific:2016aoc,LIGOScientific:2016emj}, confirming one of the most significant predictions of GR. This milestone ushered in the era of GW astronomy. However, with the deepening of our understanding of the universe, there has been an increasing number of phenomena that GR struggles to explain~\cite{Smith:1936mlg,Zwicky:1937zza,Peebles:2002gy,tHooft:1974toh,Goroff:1985th,Arkani-Hamed:1998jmv,Randall:1999ee,Randall:1999vf}. Is GR the fundamental theory of gravity? Which theory truly represents the fundamental description of gravity? These questions have become central to contemporary physics.
\par
In this paper, we investigated cosmological perturbations in the Bumblebee model with a perfect fluid; we constrained the model's parameters by requiring the absence of ghost instability, Laplacian instability, and tachyonic instability; and we further explored the properties of GWs. First, we analyzed the background field equations, incorporating the accelerated expansion of the universe. This enabled a preliminary exploration of the theory's parameter space, and a brief discussion about dark energy was given. Subsequently, we separately analyzed the scalar, vector, and tensor perturbations. To investigate stability, we eliminated non-dynamical variables from the second-order perturbed action. Under different gauge conditions, we obtained distinct effective actions containing only dynamical variables. Based on these effective actions, the stability conditions can be readily derived. By combining the stability conditions with the constraints on the speed of GWs from observational data, we constrained the parameter space of the theory in the small-scale limit. Finally, we investigated the polarization modes and propagation speeds of GWs within this constrained parameter space.
\par
Through this investigation, we reached the following conclusions. By combining the background field equations with the requirement of cosmic accelerated expansion, the time derivative of the Hubble parameter is constrained to be negative, $\dot{H}<0$~\eqref{relation-H'-1611}, and a constant $b_{t}$ implies $\sigma\ne -\tfrac{1}{2}\xi$~\eqref{vb-9242207}. The stability requirement for the vector perturbations indicates that the coupling parameter must be non-positive, $\xi\le 0$~\eqref{Tachyon-vector-1736}. Combining these two results yields constraints on dark energy: $w_{D}\le -1$ if $\sigma$ vanishes~\eqref{wD-1191247}. The stability conditions of the scalar perturbations show that the absence of ghost instability can be frequency-dependent~\eqref{ghost5-scalar-1161630}. To ensure the Bumblebee model is ghost-free at all perturbation frequencies, the ghost-free condition must reduce to the form given in Eq.~\eqref{ghost-scalar-1282053} (which, under the assumption $|\xi b_{t}^2|,|\sigma b_{t}^2|\ll 1$, requires $b_{t}^2\lesssim -\tfrac{2H^2}{\kappa\dot{H}}$). Based on the analysis combining the tensor, vector, and scalar perturbations, we found that there are seven perturbative degrees of freedom in the cosmological background, comprising two
tensor, two vector, and two scalar modes, plus one in the matter sector. To analyze the stability and propagation characteristics of GWs, we considered the small-scale limit. Based on the constraint on the speed of tensor modes derived from the GW170817 event and its electromagnetic counterpart GRB170817A, the Lorentz-violating parameter $\xi b^2_{t}$ is constrained to the range $-6\times 10^{-15} \lesssim \xi b^2_{t} \lesssim 0$~\eqref{xibt-9112245}. For the scalar modes in the small-scale limit, the Laplacian stability requires $\overline{p}_{m,n}>0$~\eqref{cm2-2137}. Furthermore, we investigated the polarization modes of GWs in the Minkowski limits. The results indicate that within the Bumblebee model, there exist two independent tensor modes ($P_{+}$ and $P_{\times}$), two independent vector modes ($P_{x}$ and $P_{y}$), and one scalar mode (the $P_{b}$ or the mixture of the $P_{b}$ and $P_{l}$), see Table \ref{Table-polarization-1108}. The propagation speed of tensor modes is less than the speed of light when $\xi b_{t}^2\ne 0$~\eqref{tensor-mode-1191254}, while the speeds of vector~\eqref{vector-mode-9261534} and scalar~\eqref{scalar-equation-lambda-2243} modes exceed it. 
\par
Due to the typically vast distances to GW sources, any deviation in the propagation speed of GWs could result in significant differences in their arrival times. This effect would cause some faint yet significant GW signals to fall outside the windows of data analysis and detection in clear signals~\cite{Schumacher:2023jxq}. Therefore, to test the Bumblebee model, it is crucial to determine the magnitude of its predicted deviations with greater precision and to expand the scope of data analysis. More importantly, detectors with higher sensitivity are necessary to increase the probability of detecting GWs with additional polarization modes. Future space-based GW detectors (Taiji, TianQin, and Lisa projects~\cite{Luo:2021qji,TianQin:2015yph,LISA:2017pwj}) are anticipated to accomplish this task.
\\

\par
\textbf{Note:} At the time of our initial submission, a related work \cite{vandeBruck:2025aaa} had already been posted on arXiv (arXiv:2509.11647). While both studies explore similar questions, there are several notable differences. The model in Ref.~\cite{vandeBruck:2025aaa} includes an additional coupling term $B_{\mu}B^{\nu}R$ (Following the reviewer's suggestion, the coupling term has been added to our latest revision.) but does not consider a separate cosmological constant term $\Lambda$. Instead, the potential term $V$ plays the role of a cosmological constant. Additionally, the authors consider a time-dependent background for the Bumblebee field, while we treat it as constant. The analysis in Ref.~\cite{vandeBruck:2025aaa} offers a detailed treatment of the dark energy implications, whereas our study focuses on the connection between perturbations and polarization modes of GWs.

\section*{Acknowledgments}
We would like to thank Shan-Ping Wu and Chen Tan for useful discussions. 
This work is supported in part by the National Key Research and Development Program of China (Grant No. 2021YFC2203003), 
the National Natural Science Foundation of China (Grants No. 12475056, No. 123B2074, and No. 12247101), Gansu Province's Top Leading Talent Support Plan, the Fundamental Research Funds for the Central Universities (Grant No. lzujbky-2025-jdzx07), the Natural Science Foundation of Gansu Province (No. 22JR5RA389 and No. 25JRRA799), the 111 Project (Grant No. B20063), and the Department of Education of Gansu Province: Outstanding Graduate ``Innovation Star" Project (Grant No. 2025CXZX-153).\par

\appendix

\section{The specific forms of some quantities} \label{appendix-1-9122141}
In this appendix, we present the specific forms of the complex quantities used in the main text.
\par
The specific forms of some quantities in the scalar perturbation equations~\eqref{equation-scalar-phil-2147} are as follows:
\begin{eqnarray}
  Q_{\phi_{h}}&=& \frac{3}{2}(\xi+2\sigma) b_{t}^2 a^3 \ddot{E} +3(1+3\sigma b_{t}^2)a^3H \dot{E} -(1+\sigma b_{t}^2)a\partial^2E -3(1-(\xi+\sigma) b_{t}^2)a^3\dot{H} E +\frac{(\xi+2\sigma) b_{t}^2}{2}a^3\partial^2\ddot{\alpha} \nonumber\\
   && +(1+3\sigma b_{t}^2)a^3H\partial^2\dot{\alpha} -(1-(\xi+\sigma) b_{t}^2)a^3\dot{H}\partial^2\alpha -(\xi+2\sigma) b_{t}^2a\partial^2\dot{\varphi}_{h} -2(1-(\xi-\sigma) b_{t}^2)a H\partial^2\varphi_{h} \nonumber\\
   && -3(\xi+2\sigma) b_{t}a^3H\dot{\phi}_{b} +(\xi+2\sigma) b_{t}a\partial^2\phi_{b} +\big(-3\xi (H^2-\dot{H})+6\sigma(H^2+\dot{H})+4\kappa b_{t}^2\overline{V}_{,b^2b^2}\big)b_{t}a^3\phi_{b} \nonumber\\
   && -2(\xi+2\sigma) b_{t}^2a\partial^2\phi_{h} -\Big(3\big(2-(\xi-8\sigma)b_{t}^2\big)H^2+9(\xi+2\sigma)b_{t}^2\dot{H}+4\kappa b_{t}^4\overline{V}_{,b^2b^2}\Big)a^3\phi_{h}\nonumber\\
   && +2\xi b_{t}aH\partial^2\varphi_{b} -\kappa\overline{\rho}_{m,n}\phi_{m}  ,\label{equation-scalar-phih-2147}
  \\
  Q_{\varphi_{h}}&=& \partial^2 \Big( (\xi+2\sigma) b_{t}^2\dot{\phi}_{h} -(2-3\xi b_{t}^2)H\phi_{h} +(1-(\xi+\sigma) b_{t}^2)\dot{E} -(\xi+2\sigma) b_{t}\dot{\phi}_{b} -(\xi-2\sigma) b_{t}H\phi_{b}\nonumber\\
   && +2(1-(\xi+\sigma) b_{t}^2)\dot{H}\varphi_{h} -\frac{\kappa\overline{\rho}_{m,n}}{a^3}\varphi_{m} \Big) ,
  \\
  Q_{E}&=& 3(\xi+2\sigma) b_{t}^2\ddot{\phi}_{h} -6\big(1-3(\xi+\sigma) b_{t}^2\big)H\dot{\phi}_{h}-\frac{2(1+\sigma b_{t}^2)}{a^2}\partial^2\phi_{h} -3\Big(3(2-3\xi b_{t}^2)H^2+(4-(5\xi+2\sigma) b_{t}^2)\dot{H} \Big)\phi_{h}\nonumber\\
   && +3(1-(\xi+\sigma) b_{t}^2)\ddot{E} +9(1-(\xi+\sigma) b_{t}^2)H\dot{E} -\frac{(1-\sigma b_{t}^2)}{a^2}\partial^2E -\frac{9\kappa\overline{J}^2\overline{\rho}_{m,nn}}{2 a^6}E +\frac{3\kappa\overline{J}\overline{\rho}_{m,nn}}{a^6}\phi_{m} \nonumber\\
   && +\partial^2\Big( (1-(\xi+\sigma) b_{t}^2)\ddot{\alpha} +3(1-(\xi+\sigma) b_{t}^2)H\dot{\alpha} -\frac{3\kappa\overline{J}^2\overline{\rho}_{m,nn}}{2 a^6}\alpha \Big) -3(\xi+2\sigma) b_{t}\ddot{\phi}_{b} -12(\xi+\sigma) b_{t}H\dot{\phi}_{b} \nonumber\\
   &&  -3(\xi-2\sigma) b_{t}(3H^2+\dot{H})\phi_{b} +\frac{4\sigma b_{t}}{a^2}\partial^{2}\phi_{b}  -\frac{2}{a^2}\partial^2 \big( (1-(\xi+\sigma) b_{t}^2)\dot{\varphi}_{h} +(1-(\xi+\sigma) b_{t}^2)H\varphi_{h} \big)\nonumber\\
   && +\frac{2\xi b_{t}}{a^2}\partial^2 \big( \dot{\varphi}_{b} +H\varphi_{b} \big) ,
  \\
  Q_{\alpha}&=& \partial^2\Big( (\xi+2\sigma) b_{t}^2\ddot{\phi}_{h} -2(1-3(\xi+\sigma) b_{t}^2)H\dot{\phi}_{h} -\big(3(2-3\xi b_{t}^2)H^2+(4-(5\xi+2\sigma) b_{t}^2)\dot{H} \big)\phi_{h} -\frac{3\kappa\overline{J}^2\overline{\rho}_{m,nn}}{2 a^6}E  \nonumber\\ 
   && +\big(1-(\xi+\sigma) b_{t}^2\big)(\ddot{E} +3H \dot{E}) -(\xi+2\sigma) b_{t}\ddot{\phi}_{b} -4(\xi+\sigma) b_{t}H\dot{\phi}_{b} -(\xi-2\sigma) b_{t}(3H^2+\dot{H})\phi_{b}\nonumber\\
   && +\frac{\kappa\overline{J}\overline{\rho}_{m,nn}}{a^6}\phi_{m} -\frac{\kappa\overline{J}^2\overline{\rho}_{m,nn}}{2 a^6}\partial^2\alpha \Big)  ,
  \\
  Q_{\phi_{b}}&=& -3(\xi+2\sigma) b_{t}\ddot{E}-6(\xi+4\sigma) b_{t}H \dot{E} +\frac{4\sigma b_{t}}{a^2}\partial^2E -(\xi+2\sigma) b_{t}\partial^2\ddot{\alpha}-2(\xi+4\sigma) b_{t}H\partial^2\dot{\alpha} +6(\xi+2\sigma) b_{t}H\dot{\phi}_{h} \nonumber\\
   && +\frac{2(\xi+2\sigma) b_{t}}{a^2}\partial^2\phi_{h} +4b_{t}\big(3(\xi+4\sigma)H^2+3(\xi+2\sigma)\dot{H}+2\kappa b_{t}^2\overline{V}_{b^2b^2} \big)\phi_{h} +\frac{2(\xi+2\sigma) b_{t}}{a^2}\partial^2\dot{\varphi}_{h} +\frac{8\sigma b_{t}}{a^2}H\partial^2\varphi_{h} \nonumber\\
   && +\frac{2\kappa}{a^2}\partial^2\dot{\varphi}_{b} -\frac{2\kappa}{a^2}\partial^2\phi_{b} -8\kappa b_{t}^2\overline{V}_{,b^2b^2}\phi_{b} ,
  \\
  Q_{\varphi_{b}}&=& \partial^2\Big( \ddot{\varphi}_{b} +H\dot{\varphi}_{b} +\frac{2\xi}{\kappa}\dot{H}\varphi_{b} -\frac{\xi b_{t}}{\kappa}\dot{E} -\dot{\phi}_{b} -H\phi_{b} +\frac{2\xi b_{t}}{\kappa}H\phi_{h} \Big) ,
  \\
  Q_{\phi_{m}}&=& \dot{\phi}_{\ell} +\overline{\rho}_{m,n}\phi_{h} -\frac{\overline{J}\overline{\rho}_{m,nn}}{2 a^3}(3E+\partial^2\alpha) +\frac{\overline{\rho}_{m,nn}}{a^3}\phi_{m}  ,
  \\
  Q_{\varphi_{m}}&=& \partial^2\Big( \overline{\rho}_{m,n}\varphi_{h} -\phi_{\ell} +\frac{\overline{\rho}_{m,n}}{\overline{J}}\varphi_{m} \Big)  ,
  \\
  Q_{\phi_{\ell}}&=& a^2\dot{\phi}_{m} +\partial^2\varphi_{m}.
\end{eqnarray}

\bibliographystyle{unsrt}
\bibliography{referenceData}

\end{document}